\PassOptionsToPackage{unicode}{hyperref}
\PassOptionsToPackage{hyphens}{url}
\PassOptionsToPackage{dvipsnames,svgnames,x11names}{xcolor}
\documentclass[12pt]{article}

\usepackage{amsmath,amssymb}
\usepackage{amsthm,bm}
\usepackage{iftex}
\ifPDFTeX
  \usepackage[T1]{fontenc}
  \usepackage[utf8]{inputenc}
  \usepackage{textcomp} % provide euro and other symbols
\else % if luatex or xetex
  \usepackage{unicode-math}
  \defaultfontfeatures{Scale=MatchLowercase}
  \defaultfontfeatures[\rmfamily]{Ligatures=TeX,Scale=1}
\fi
\usepackage{lmodern}
\usepackage{tabularx}
\usepackage{setspace}
\usepackage{multirow}
\usepackage{booktabs}
\usepackage{caption}
\usepackage{algorithm}
\usepackage{algpseudocode}
\usepackage{setspace}
\usepackage{bookmark}

\ifPDFTeX\else  
\fi
\IfFileExists{upquote.sty}{\usepackage{upquote}}{}
\IfFileExists{microtype.sty}{% use microtype if available
  \usepackage[]{microtype}
  \UseMicrotypeSet[protrusion]{basicmath} % disable protrusion for tt fonts
}{}
\makeatletter
\@ifundefined{KOMAClassName}{% if non-KOMA class
  \IfFileExists{parskip.sty}{%
    \usepackage{parskip}
  }{% else
    \setlength{\parindent}{0pt}
    \setlength{\parskip}{6pt plus 2pt minus 1pt}}
}{% if KOMA class
  \KOMAoptions{parskip=half}}
\makeatother
\usepackage{xcolor}
\makeatletter
\ifx\paragraph\undefined\else
  \let\oldparagraph\paragraph
  \renewcommand{\paragraph}{
    \@ifstar
      \xxxParagraphStar
      \xxxParagraphNoStar
  }
  \newcommand{\xxxParagraphStar}[1]{\oldparagraph*{#1}\mbox{}}
  \newcommand{\xxxParagraphNoStar}[1]{\oldparagraph{#1}\mbox{}}
\fi
\ifx\subparagraph\undefined\else
  \let\oldsubparagraph\subparagraph
  \renewcommand{\subparagraph}{
    \@ifstar
      \xxxSubParagraphStar
      \xxxSubParagraphNoStar
  }
  \newcommand{\xxxSubParagraphStar}[1]{\oldsubparagraph*{#1}\mbox{}}
  \newcommand{\xxxSubParagraphNoStar}[1]{\oldsubparagraph{#1}\mbox{}}
\fi
\makeatother

\providecommand{\trans}{\mathsf T}

\usepackage{longtable,booktabs,array}
\usepackage{calc} % for calculating minipage widths
\usepackage{etoolbox}
\makeatletter
\patchcmd\longtable{\par}{\if@noskipsec\mbox{}\fi\par}{}{}
\makeatother
\IfFileExists{footnotehyper.sty}{\usepackage{footnotehyper}}{\usepackage{footnote}}
\makesavenoteenv{longtable}
\usepackage{graphicx}
\makeatletter
\def\maxwidth{\ifdim\Gin@nat@width>\linewidth\linewidth\else\Gin@nat@width\fi}
\def\maxheight{\ifdim\Gin@nat@height>\textheight\textheight\else\Gin@nat@height\fi}
\makeatother
\setkeys{Gin}{width=\maxwidth,height=\maxheight,keepaspectratio}
\makeatletter
\def\fps@figure{htbp}
\makeatother

\makeatletter
\@ifpackageloaded{caption}{}{\usepackage{caption}}
\AtBeginDocument{%
\ifdefined\contentsname
  \renewcommand*\contentsname{Table of contents}
\else
  \newcommand\contentsname{Table of contents}
\fi
\ifdefined\listfigurename
  \renewcommand*\listfigurename{List of Figures}
\else
  \newcommand\listfigurename{List of Figures}
\fi
\ifdefined\listtablename
  \renewcommand*\listtablename{List of Tables}
\else
  \newcommand\listtablename{List of Tables}
\fi
\ifdefined\figurename
  \renewcommand*\figurename{Figure}
\else
  \newcommand\figurename{Figure}
\fi
\ifdefined\tablename
  \renewcommand*\tablename{Table}
\else
  \newcommand\tablename{Table}
\fi
}

\newtheorem{proposition}{Proposition}
\newtheorem{assumption}{Assumption}
\newtheorem{theorem}{Theorem}

\newtheorem{corollary}{Corollary}
\newtheorem{rem}{Remark}

\@ifpackageloaded{float}{}{\usepackage{float}}
\floatstyle{ruled}
\@ifundefined{c@chapter}{\newfloat{codelisting}{h}{lop}}{\newfloat{codelisting}{h}{lop}[chapter]}
\floatname{codelisting}{Listing}

\makeatother
\makeatletter
\@ifpackageloaded{caption}{}{\usepackage{caption}}
\@ifpackageloaded{subcaption}{}{\usepackage{subcaption}}
\makeatother

\ifLuaTeX
  \usepackage{selnolig}  % disable illegal ligatures
\fi
\usepackage[]{natbib}
\setcitestyle{aysep={},citesep={;}}
\usepackage{bookmark}

\IfFileExists{xurl.sty}{\usepackage{xurl}}{} % add URL line breaks if available
\hypersetup{
  pdftitle={Cross-Order Information-Guided Estimation for Non-Uniform Hypergraphs},
  pdfauthor={Rongzhuo Nie; Menghao Wang; Sanguo Zhang; Mingyang Ren; Guojun Zhu},
  pdfkeywords={Non-uniform hypergraphs; Latent space models; Computational efficiency; Negative sampling; Cross-order information.},
  colorlinks=true,
  linkcolor={blue},
  filecolor={Maroon},
  citecolor={Blue},
  urlcolor={Blue},
  pdfcreator={LaTeX via pandoc}}

\graphicspath{{../}}
\ifPDFTeX\pdfminorversion=7\fi
\hypersetup{hypertexnames=false}
\theoremstyle{remark}

\date{}

\newcommand{\anon}{1}

\usepackage{pdfpages}
\begin{document}

\def\spacingset#1{\renewcommand{\baselinestretch}%
{#1}\small\normalsize} \spacingset{1}

%%%%%%%%%%%%%%%%%%%%%%%%%%%%%%%%%%%%%%%%%%%%%%%%%%%%%%%%%%%%%%%%%%%%%%%%%%%%%%

\if1\anon
{
  \title{\Large\bf Cross-Order Information-Guided\\Estimation for Non-Uniform Hypergraphs}
  \author{\parbox{\textwidth}{\centering\normalsize
Rongzhuo Nie\textsuperscript{1,2}\thanks{Rongzhuo Nie and Menghao Wang contributed equally to this work.},\enspace
Menghao Wang\textsuperscript{1,2}\begin{NoHyper}\footnotemark[1]\end{NoHyper},\enspace
Sanguo Zhang\textsuperscript{1,2},\\
Mingyang Ren\textsuperscript{3}\thanks{Corresponding authors. Emails:
\url{renmingyang17@mails.ucas.ac.cn};
\url{zhuguojun23@mails.ucas.ac.cn}.},\enspace
Guojun Zhu\textsuperscript{1,2}\begin{NoHyper}\footnotemark[2]\end{NoHyper}
\\[0.5em]
% \fontsize{9}{11}\selectfont
\textsuperscript{1}School of Mathematical Sciences, University of Chinese Academy of Sciences, \\Beijing, China\\
\textsuperscript{2}Key Laboratory of Big Data Mining and Knowledge Management, \\Chinese Academy of Sciences, Beijing, China\\
\textsuperscript{3}School of Mathematical Sciences, Shanghai Jiao Tong University, Shanghai, China
}}
  \setlength{\tabcolsep}{0pt} % Local to the title block.
  \maketitle
} \fi

\if0\anon
{
  \bigskip
  \bigskip
  \bigskip
  \begin{center}
    {\Large\bf Cross-Order Information-Guided\\Estimation for Non-Uniform Hypergraphs}
\end{center}
  \medskip
} \fi

\bigskip
\begin{abstract}
Many real-world interactions involve groups of nodes rather than pairs. A non-uniform hypergraph records such interactions as hyperedges of different orders, where the order counts the nodes in a hyperedge. Full estimation evaluates every candidate hyperedge, whose number grows exponentially with the order, and classical remedies sacrifice higher-order information or aggravate computation. The orders are not isolated: the lower orders, cheap to compute and rich in information, reveal where higher-order hyperedges are likely to occur. We propose {\bf CINE} (Cross-order Information-guided Negative-sampling Estimation) under a latent space model in which all orders share latent positions. Starting from order two, CINE climbs one order at a time. At each higher order, CINE keeps all present hyperedges, samples a small subset of negative examples guided by lower-order scores, and corrects for the known inclusion probabilities to preserve the target of full estimation. CINE thereby lets users choose the trade-off between computation time and estimation accuracy: the retained orders and the sampling budget set the time, and the retained information provably approaches full-data information as the sampling budget grows. In simulations, CINE approaches the accuracy of full estimation at a small fraction of its computation time, and larger teams in MADStat reveal research associations.
\end{abstract}

\noindent%
\vfill

\newpage
\spacingset{1.8} % DON'T change the spacing!

\section{Introduction}\label{Section-Introduction}

Over the past decades, the statistical analysis of network data has focused mainly on pairwise interactions, that is, on whether an edge connects two nodes \citep{hoff2002latent,fienberg2012brief}. In the real world, however, many interactions occur among a group of nodes at once: the coauthors of one paper, the participants of one meeting, the molecules of one chemical reaction, and the items on one shopping receipt \citep{battiston2020networks}. Breaking such a group interaction into pairwise edges loses the information on which nodes interact together: one paper by the four authors in \(\{a,b,c,d\}\) and six papers by the six author pairs \(\{a,b\},\{a,c\},\ldots,\{c,d\}\) produce exactly the same pairwise edges \citep{turnbull2024latent}. The hypergraph retains this information. In a hypergraph, a hyperedge can connect any number of nodes, and the number of nodes in a hyperedge is its order, so an ordinary edge is a hyperedge of order two. Since groups in practice differ in size, one hypergraph often contains hyperedges of several orders, and we call such a hypergraph non-uniform \citep{zhen2023community}. Statistical modeling of non-uniform hypergraphs has thus become an important direction of network analysis, with applications to scientific collaboration and biological networks \citep{contisciani2022inference,turnbull2024latent}.

\begin{figure}[ht]
    \centering
    \includegraphics[width=\linewidth]{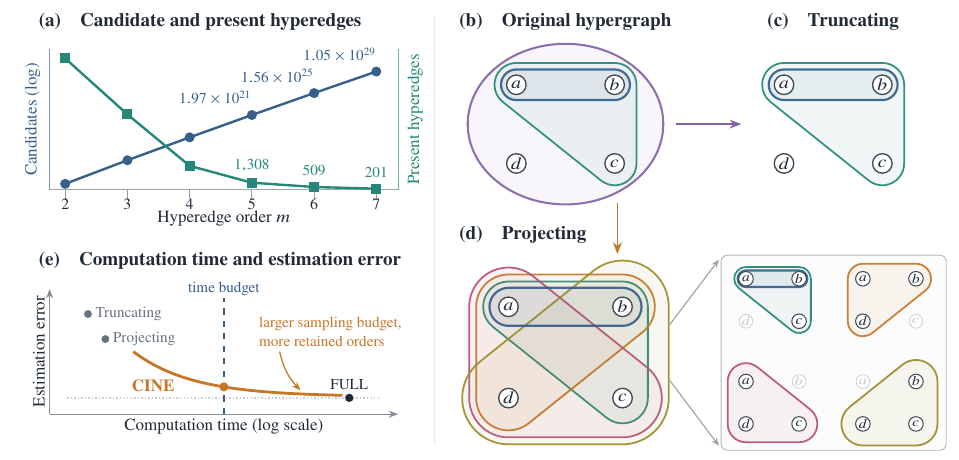}
    \caption{Growth in the number of candidate hyperedges and compression of non-uniform hypergraphs. (a) Numbers of candidate hyperedges and present hyperedges across orders in MADStat. (b) An original hypergraph containing $\{a,b\}$, $\{a,b,c\}$, and $\{a,b,c,d\}$. Each colored region encloses the members of one hyperedge. (c) Truncating at order three discards $\{a,b,c,d\}$. (d) Projecting replaces $\{a,b,c,d\}$ with its three-node subsets while retaining the existing lower-order hyperedges. The right panel shows the resulting hyperedges. (e) Schematic trade-off between computation time and estimation error. Truncating and Projecting are fast but lose higher-order information, FULL is accurate but slow, and CINE traces a curve between them whose position the user selects through the retained orders and the sampling budget.}
    \label{fig:motivation}
\end{figure}

To estimate the generating mechanism of a non-uniform hypergraph, a common class of statistical models evaluates every candidate hyperedge in the objective function, such as the negative log-likelihood, and asks whether the candidate hyperedge is present \citep{10.1214/25-AOS2524}. Here a candidate hyperedge of order \(m\) is any set of \(m\) distinct nodes. The number of candidate hyperedges grows explosively with the order, while present hyperedges occupy only a tiny fraction of these candidate hyperedges. For example, as Figure~\hyperref[fig:motivation]{1(a)} illustrates, in the MADStat data \citep{ji2022co}, the number of candidate hyperedges increases from approximately $1.97\times10^{21}$ at order five to $1.05\times10^{29}$ at order seven, whereas the number of present hyperedges decreases from $1{,}308$ to $201$. The computation time of full estimation, which evaluates all candidate hyperedges, therefore grows exponentially with the order, and full estimation of all orders is almost impossible for a moderately large hypergraph. Existing works thus usually compress the hypergraph before estimation, most classically by Truncating or Projecting. As Figure~\hyperref[fig:motivation]{1(b)}--\hyperref[fig:motivation]{1(d)} illustrates, Truncating discards all hyperedges above a chosen order \citep{zhen2023community, lucas2026reducibility, zhu2026fold}, whereas Projecting breaks a higher-order hyperedge into hyperedges of a lower order \citep{agarwal2006higher, zhou2006learning, chodrow2021generative}. Methods for uniform hypergraphs, in which all hyperedges share the same order, are also abundant \citep{ghoshdastidar2017uniform,lyu2023latent}. To apply these methods, many works first adopt Padding, which pads every lower-order hyperedge up to the highest order with a null node, an artificial node with no real counterpart. Since the highest order is often still too high to handle, Truncating or Projecting is then applied \citep{zhen2023community, zhu2026fold}. Truncating throws away the information in higher-order hyperedges, and Projecting blends higher-order information into lower orders and overweights large hyperedges: one paper by seven authors becomes 35 three-author collaborations, which pull the estimation toward the seven authors \citep{yoon2020much, contisciani2022inference}. Padding loses no information, but Padding lifts every order to the highest order and adds an artificial node, so the computation grows rather than shrinks. The estimation of non-uniform hypergraphs thus faces a dilemma: either spend enormous computation time on all orders or sacrifice higher-order information. Ideally, the analyst would set the computation time that can be afforded and obtain an estimation accuracy that approaches that of full estimation as more time is allowed. Reaching this goal requires a method that does not evaluate every higher-order candidate hyperedge, yet retains every higher-order hyperedge.

Beyond Truncating, Projecting, and Padding, some works handle non-uniform hypergraphs without compressing the data \citep{turnbull2024latent,yu2025modeling}. One line of work evaluates only a random subset of candidate hyperedges and corrects the likelihood for the sampling, through case-control sampling \citep{raftery2012fast,lerner2023relational} or negative sampling \citep{shi2023surprising}, an idea closely related to noise-contrastive estimation \citep{gutmann2012nce}. Since the vast majority of candidate hyperedges in a uniformly drawn subset are not present and contribute almost nothing to estimation, these works must evaluate many candidate hyperedges to preserve estimation accuracy, and the computation time rises again. Local case-control sampling \citep{fithian2014local} draws more informative samples with a pilot fit, but evaluating the pilot fit on every target-order candidate hyperedge is exactly the computation that explodes. Another line exploits sparsity and organizes the computation around the present hyperedges: \citet{yu2025modeling} sum likelihood contributions over present hyperedges and compute the normalizing term analytically, and \citet{turnbull2024latent} evaluate the likelihood using overlap counts between present and geometrically induced hyperedges. Ultimately, each of these works accelerates the estimation of one order from within that order alone.

Yet all of these works overlook the simplest fact: the orders of a non-uniform hypergraph are not isolated islands. The same nodes appear across orders. For example, in MADStat \citep{ji2022co}, 2,421 of the 6,616 hyperedges of order four or higher contain a pair or triple that had already formed a lower-order hyperedge in an earlier year. Moreover, information and computation are misaligned across orders. A lower order has few candidate hyperedges and many present hyperedges, so computation is cheap and information is rich. A higher order has many candidate hyperedges and few present hyperedges, so computation is expensive and information is sparse. The lower orders therefore reveal which higher-order candidate hyperedges deserve evaluation, and we call the information that lower orders provide about higher-order hyperedges cross-order information. We start from order two, where computation is cheapest, and climb one order at a time. At each new order, cross-order information, which need not come from the adjacent order alone, guides the choice of candidate hyperedges to evaluate. In this way, the number of candidate hyperedges to evaluate no longer explodes with the order, and every higher-order hyperedge remains in the estimation, which preserves the estimation accuracy.

Statistically, every candidate hyperedge corresponds to a binary observation, which equals one if the candidate hyperedge is present and zero otherwise. For such data, the latent space model \citep{hoff2002latent,turnbull2024latent} represents each node by a latent position, and a group of nodes with closer latent positions is more likely to form a hyperedge. The latent space model has been extended to various hypergraph data: \citet{yuan2023high} jointly model pairwise and higher-order links, \citet{zhen2023community} embed non-uniform hypergraphs after null-node padding, and \citet{Wu27082026} model sparse non-uniform hypergraphs with repeated hyperedges. In this paper, we let all orders share one set of latent positions, so the latent positions estimated at lower orders can score every higher-order candidate hyperedge directly and carry cross-order information. On this basis, we propose CINE (Cross-order Information-guided Negative-sampling Estimation). We call a candidate hyperedge that is not present a negative example. At each higher order, CINE keeps all present hyperedges, evaluates only a small subset of negative examples that CINE selects according to scores from the lower orders, and corrects the estimation with the known inclusion probabilities, that is, the probabilities that the negative examples are sampled, so the estimation target of CINE coincides with the target of full estimation (Section~\ref{Section-Methodology}). The computation time of CINE thus grows with the numbers of present hyperedges and sampled negative examples rather than with the number of target-order candidate hyperedges, while the estimation accuracy of CINE stays close to that of full estimation. The trade-off between computation time and estimation accuracy thus becomes a choice of the user, as Figure~\hyperref[fig:motivation]{1(e)} illustrates: given a time budget, the retained orders are chosen within the budget (Remark~\ref{rem:order-selection}), and a larger sampling budget brings the estimation accuracy closer to that of full estimation (Section~\ref{Section-Theory}).

Our contributions are fourfold. (i) We identify the dilemma between computation time and estimation accuracy in the estimation of non-uniform hypergraphs and show that cross-order information turns the dilemma into a trade-off chosen by the user: CINE keeps every present hyperedge under any time budget and approaches full estimation as the time budget grows. (ii) We propose a latent space model in which all orders share latent positions, together with CINE, an estimation method that climbs order by order under the guidance of cross-order information. (iii) We establish that CINE retains most of the full-data information and derive nonasymptotic error bounds in which the latent positions attain the rate of the most informative retained order and the effect of the sampling budget is explicit, showing that CINE attains estimation accuracy comparable to full estimation at substantially lower computational cost. (iv) In simulations, CINE attains lower estimation errors than the methods of \citet{yuan2023high}, \citet{zhen2023community}, and \citet{Wu27082026}, at a small fraction of the computation time of its main competitors, and the accuracy of CINE improves steadily as the sampling budget grows. On the MADStat data \citep{ji2022co}, incorporating larger teams through CINE reveals associations among statisticians with related research interests.

\section{Methodology}
\label{Section-Methodology}

\subsection{Shared Latent Space Model for Non-uniform Hypergraphs}

A hypergraph is denoted by $G=(V,E)$, where $V=\{v_1,\ldots,v_n\}$ is a set of $n$ nodes and each hyperedge $e\in E$ is an unordered subset of $V$ whose order is its cardinality. We focus on non-uniform hypergraphs, in which hyperedges may differ in order. The presence of a hyperedge does not require its proper subsets to be present as hyperedges, which suits data recording joint participation in events: an event involving a given group and one involving only a subset of its members are distinct observations.

Let $\mathcal{H}$ denote the set of hyperedge orders present in $G$. Including an order adds computational cost, and some orders contain very few observed hyperedges. We therefore fit the model on a set of retained hyperedge orders $\mathcal{M}\subseteq\mathcal{H}$, which can be chosen under a time budget as in Remark~\ref{rem:order-selection}. We define $\Omega_m=\{e\subseteq[n]:|e|=m\}$ as the set of all possible hyperedges of order $m$, termed order-$m$ candidate hyperedges (candidates for short), where $[n]=\{1,2,\ldots,n\}$. The binary indicator $A_e^{(m)}=\mathbf{1}\{e\in E\}$ records whether candidate $e$ is present in $G$, where $\mathbf{1}\{\cdot\}$ denotes the indicator function. Candidates with $A_e^{(m)}=1$ are present hyperedges, and candidates with $A_e^{(m)}=0$ are referred to as negative examples. We denote by $E_m^+=\{e\in\Omega_m:A_e^{(m)}=1\}$ the set of order-$m$ present hyperedges. Real-world hypergraphs are typically sparse: only a small fraction of candidate hyperedges is present, and we focus on such sparse hypergraphs.

To parameterize the occurrence probability of each candidate hyperedge, we adopt latent space modeling \citep{hoff2002latent,ma2020universal}, which assigns each node a latent position.
Specifically, each node $v_i$ is associated with a latent position $\bm{z}_i\in\mathbb{R}^r$, where $r$ is the latent dimension. Collecting these vectors as rows gives the latent position matrix $\bm{Z}\in\mathbb{R}^{n\times r}$, which is common to all orders in $\mathcal{M}$.

For a candidate hyperedge $e\in\Omega_m$, we assume that the occurrence probability $p_m(e;\bm{Z},\beta_m)$ depends only on the shared latent position matrix $\bm{Z}$ and an order-specific intercept $\beta_m$ through the logistic model:
\begin{equation}\label{eq:model}
    p_m(e;\bm{Z},\beta_m):=\operatorname{logit}^{-1}(\eta_m(e; \bm{Z},\beta_m))\quad \text{and} \quad\eta_m(e; \bm{Z},\beta_m):=\beta_m+s_m(e;\bm{Z}),
\end{equation}
where $\operatorname{logit}(x)=\log\{x/(1-x)\}$ and the score $s_m(e;\bm{Z})$, which summarizes the aggregate latent affinity among the nodes of $e$, is:
\begin{equation}\label{eq:score}
    s_m(e;\bm{Z}) = \left\{r\binom{m}{2}\right\}^{-1/2} \sum_{\{i,j\}\subseteq e}\bm{z}_i^\top \bm{z}_j.
\end{equation}
The normalization by $\{r\binom{m}{2}\}^{1/2}$ adjusts the scale of the score for the latent dimension and the number of node pairs. The intercept $\beta_m$ determines the log odds of occurrence when the score is zero and allows the baseline hyperedge prevalence to vary across orders. For fixed $m$ and $\beta_m$, a larger score gives a higher occurrence probability. When hyperedges represent joint participation in events, the score has a natural interpretation: just as the number of events shared by pairs of group members, summed over all pairs within the group, reflects the strength of ties among the members, the score aggregates pairwise latent affinities into the propensity of the group to participate jointly in an event.

Conditional on $\bm{Z}$ and $\beta_m$, the indicators $A_e^{(m)}$ are assumed to be mutually independent with $A_e^{(m)} \sim \operatorname{Bernoulli}\{p_m(e;\bm{Z},\beta_m)\}$. The latent positions are not unique, since every orthogonal matrix $\bm{O}\in\mathbb{R}^{r\times r}$ gives $s_m(e;\bm{Z}\bm{O})=s_m(e;\bm{Z})$ and hence $p_m(e;\bm{Z}\bm{O},\beta_m)=p_m(e;\bm{Z},\beta_m)$. We impose the centering constraint $\sum_{i=1}^n\bm{z}_i=\bm{0}$ throughout, under which the model is identifiable up to a common orthogonal transformation.

\begin{proposition}[Identifiability]\label{prop:1}
Suppose that \(2\in\mathcal M\) and \(n\ge r+2\). Under the centering constraint stated above, the probabilities \(p_m(e; \boldsymbol{Z},\beta_m)\), for all \(m\in\mathcal M\) and \(e\in\Omega_m\), uniquely determine the Gram matrix \(\boldsymbol{Z}\boldsymbol{Z}^\top\) and the order-specific intercepts \(\{\beta_m\}_{m\in\mathcal M}\). Consequently, the shared latent position matrix \(\boldsymbol{Z}\) is determined up to a common orthogonal transformation.
\end{proposition}
We therefore treat latent position matrices that differ by an orthogonal transformation as equivalent and work with the Gram matrix $\bm{Z}\bm{Z}^\top$.

\subsection{Cross-order Information-guided Estimation}

We aim to estimate $\bm{Z}$ and $\{\beta_m\}_{m\in\mathcal{M}}$. The full Bernoulli loss of $G$ is the Bernoulli negative log-likelihood over all candidates:
\begin{equation}\label{eq:full-loss}
    \begin{aligned}
        \ell\!\left(\bm{Z},\{\beta_m\}_{m\in\mathcal M}\right)&=-\sum_{m\in\mathcal M}\sum_{e\in\Omega_m}\left[A_e^{(m)}\log p_m(e;\bm{Z},\beta_m)+\{1-A_e^{(m)}\}\log\{1-p_m(e;\bm{Z},\beta_m)\}\right]\\
        &=\sum_{m\in\mathcal M}\sum_{e\in\Omega_m}\left[\log\{1+\exp\{\eta_m(e;\bm{Z},\beta_m)\}\}-A_e^{(m)}\eta_m(e;\bm{Z},\beta_m)\right].
    \end{aligned}
\end{equation}
In \eqref{eq:full-loss}, each order-$m$ negative example contributes the strictly positive term $\log\{1+\exp(\beta_m+s_m(e;\bm{Z}))\}$, so even for a sparse hypergraph, full estimation must evaluate all $\sum_{m\in\mathcal{M}}\binom{n}{m}$ candidates. Since most candidates are negative examples, we include all present hyperedges and sample only a subset of negative examples.

\begin{figure}[ht]
    \centering
    \includegraphics[width=\linewidth]{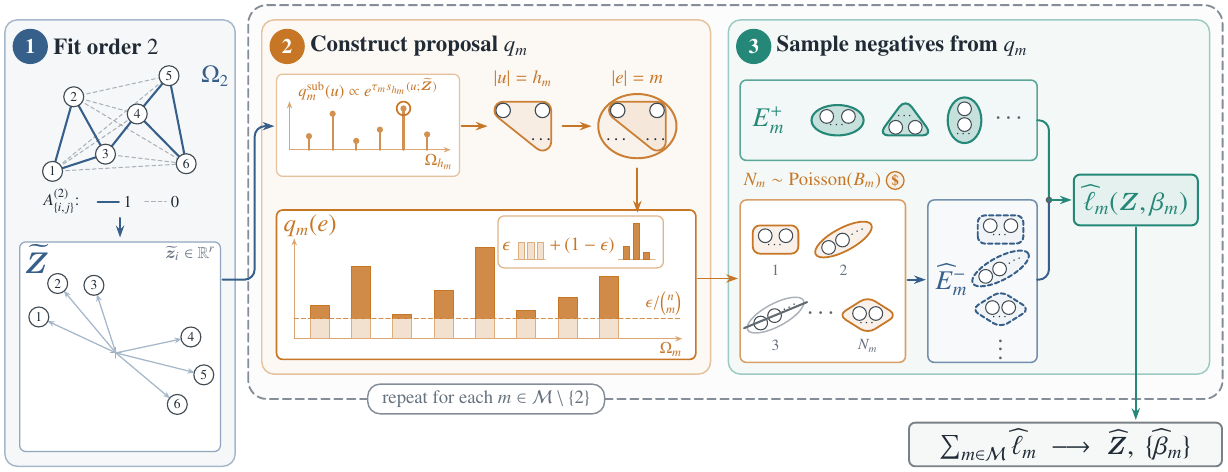}
    \caption{Workflow of CINE. (1) CINE minimizes the order-two loss over all order-two candidates $\Omega_2$ and obtains the preliminary estimate $\widetilde{\bm Z}$ of the shared latent positions. (2) For each retained order $m>2$, the preliminary scores of the candidates of order $h_m$, the largest retained order below $m$, define the selection probabilities $q_m^{\mathrm{sub}}(u)$. Completing a selected $u$ with $m-h_m$ uniformly drawn nodes and mixing with uniform sampling of weight $\epsilon$ gives the proposal $q_m$ in \eqref{eq:proposal}. (3) CINE draws $N_m\sim\operatorname{Poisson}(B_m)$ candidates from $q_m$, keeps all present hyperedges $E_m^+$ together with the sampled negative examples $\widehat E_m^-$, and forms the corrected loss $\widehat\ell_m(\bm Z,\beta_m)$ with the offsets $-\log\pi_m(e)$. Steps (2) and (3) repeat for every $m\in\mathcal M\setminus\{2\}$, and minimizing the CINE loss $\sum_{m\in\mathcal M}\widehat\ell_m$ in \eqref{eq:cine} gives $\widehat{\bm Z}$ and $\{\widehat\beta_m\}$.}
    \label{fig:flowchart}
\end{figure}

The sampling is guided by cross-order information. We require $2\in\mathcal M$ (Proposition~\ref{prop:1}) and start from order two, where computation is cheapest. First, as shown in panel~(1) of Figure~\ref{fig:flowchart}, we fit the model to the order-two candidates by minimizing the order-two loss, that is, the order-two component of the full Bernoulli loss:
\begin{equation*}
    (\widetilde{\bm{Z}},\widetilde{\beta}_2)\in\underset{\substack{\bm{Z},\beta_2\\\sum_{i=1}^n \bm{z}_i=\bm{0}}}{\operatorname{arg\,min}}\sum_{1\le i<j\le n}\left[\log\left\{1+\exp\left(\beta_2+\frac{\bm{z}_i^\top \bm{z}_j}{\sqrt{r}}\right)\right\}-A_{\{i,j\}}^{(2)}\left(\beta_2+\frac{\bm{z}_i^\top \bm{z}_j}{\sqrt{r}}\right)\right].
\end{equation*}
The preliminary estimate $\widetilde{\bm{Z}}$ of the shared latent position matrix is cheap to compute, since the order-two candidate set contains only $\binom{n}{2}=O(n^2)$ elements.

Second, as shown in panel~(2) of Figure~\ref{fig:flowchart}, for each target order $m\in\mathcal{M}$ with $m>2$, we construct a cross-order proposal distribution from $\widetilde{\bm{Z}}$. Let $h_m=\max\{k\in\mathcal{M}:k<m\}$ denote the largest retained order below $m$, which is well defined since $2\in\mathcal{M}$. For every $e\in\Omega_{h_m}$, present or not, we compute the preliminary score $s_{h_m}(e;\widetilde{\bm{Z}})=\{r\binom{h_m}{2}\}^{-1/2}\sum_{\{i,j\}\subseteq e}\widetilde{\bm{z}}_i^\top\widetilde{\bm{z}}_j$, where $\widetilde{\bm{z}}_i^\top$ denotes the $i$-th row of $\widetilde{\bm{Z}}$.

For each target order $m$, a single draw selects $e \in \Omega_m$ with probability:
\begin{equation}\label{eq:proposal}
    q_m(e)=\frac{\epsilon}{\binom{n}{m}}+\frac{1-\epsilon}{\binom{n-h_m}{m-h_m}}\sum_{\substack{u\subseteq e\\|u|=h_m}}q_m^{\mathrm{sub}}(u),
\end{equation}
where the fixed mixture weight $\epsilon\in(0,1]$ is the probability of sampling uniformly from $\Omega_m$, and $q_m^{\mathrm{sub}}(u)=\exp\{\tau_m s_{h_m}(u;\widetilde{\bm{Z}})\}/\sum_{v\in\Omega_{h_m}}\exp\{\tau_m s_{h_m}(v;\widetilde{\bm{Z}})\}$ is the probability of selecting $u\in\Omega_{h_m}$. The scaling factor $\tau_m=\{\binom{m}{2}/\binom{h_m}{2}\}^{1/2}$ accounts for differences in score normalization across orders: if the preliminary pairwise inner products within a target candidate $e$ are all equal, then $\tau_m s_{h_m}(u;\widetilde{\bm{Z}})=s_m(e;\widetilde{\bm{Z}})$ for every $u\in\Omega_{h_m}$ contained in $e$. The probabilities $q_m^{\mathrm{sub}}(u)$ favor order-$h_m$ candidate hyperedges with larger preliminary scores and depend only on $\widetilde{\bm{Z}}$ and the retained orders.

A draw from \eqref{eq:proposal} is generated as follows. With probability $\epsilon$, we draw $e$ uniformly from $\Omega_m$. With probability $1-\epsilon$, we select $u\in\Omega_{h_m}$ with probability $q_m^{\mathrm{sub}}(u)$ and complete $u$ by drawing $m-h_m$ nodes uniformly without replacement from $[n]\setminus u$. The denominator $\binom{n-h_m}{m-h_m}$ counts the possible completions of $u$, and the summation runs over all order-$h_m$ candidate hyperedges contained in $e$, so $\sum_{e\in\Omega_m}q_m(e)=1$ and $q_m(e)\ge\epsilon/\binom{n}{m}>0$ for every $e\in\Omega_m$. Compared with uniform sampling ($\epsilon=1$), the proposal distribution favors order-$m$ candidate hyperedges that contain a lower-order candidate hyperedge with a large preliminary score. Directing sampling toward node groups with stronger potential associations yields more informative negative examples under a limited sampling budget. The mixture weight $\epsilon$ controls the contribution of uniform sampling: larger values move the proposal distribution closer to uniform, whereas smaller values place greater emphasis on the preliminary scores.

Third, as shown in panel~(3) of Figure~\ref{fig:flowchart}, for each target order $m$, we generate the number of proposal draws $N_m\sim\operatorname{Poisson}(B_m)$ and draw $N_m$ candidates independently with replacement from $q_m$. Let $C_e^{(m)}$ denote the number of times $e\in\Omega_m$ is drawn, so that $C_e^{(m)}\mid N_m=k\sim\operatorname{Binomial}\{k,q_m(e)\}$. By the law of total probability,
\begin{equation*}
    \begin{aligned}
        \Pr\{C_e^{(m)}=0\}&=\sum_{k=0}^{\infty}\Pr\{N_m=k\}\Pr\{C_e^{(m)}=0\mid N_m=k\}\\
        &=\sum_{k=0}^{\infty}\exp(-B_m)\frac{B_m^k}{k!}\{1-q_m(e)\}^k=\exp\{-B_mq_m(e)\}.
    \end{aligned}
\end{equation*}
The inclusion probability of $e$, namely the probability that $e$ is drawn at least once, is therefore $\pi_m(e)=1-\exp\{-B_mq_m(e)\}$. Removing duplicates and present hyperedges from the draws yields the set of sampled negative examples $\widehat E_m^-=\{e\in\Omega_m:A_e^{(m)}=0,\ C_e^{(m)}\ge1\}$, whose expected size is at most $B_m$. We therefore call $B_m$ the \textit{sampling budget} for order $m$. We set $B_m$ through the budget ratio $B_m/\mu_m$, or $B_m/|E_m^+|$ when $\mu_m$ is unknown, which controls the fraction of full-data information retained (Section~\ref{subsec:sampling}).

To account for the unequal inclusion probabilities induced by negative sampling, we construct a conditional Bernoulli likelihood for the included candidates. For each target order $m>2$ and candidate $e\in\Omega_m$, let $R_e^m=\mathbf{1}\{e\in E_m^+\cup\widehat E_m^-\}$ indicate whether $e$ is included in the estimation sample. Throughout the following derivation, we condition on the preliminary estimate and regard the proposal probabilities $q_m(e)$ and sampling budget $B_m$ as fixed. Since every present hyperedge is included, whereas a negative example is included with probability $\pi_m(e)$, we have $\Pr\{R_e^m=1\mid A_e^{(m)}=1\}=1$ and $\Pr\{R_e^m=1\mid A_e^{(m)}=0\}=\pi_m(e)$. Bayes' rule and the logistic model therefore give:
\begin{equation*}
    \begin{aligned}
        \Pr\{A_e^{(m)}=1\mid R_e^m=1\}&=\frac{\Pr\{A_e^{(m)}=1,R_e^m=1\}}{\Pr\{R_e^m=1\}}\\
        &=\frac{p_m(e;\bm{Z},\beta_m)}{p_m(e;\bm{Z},\beta_m)+\{1-p_m(e;\bm{Z},\beta_m)\}\pi_m(e)}\\
        &=\frac{\exp\{\eta_m(e;\bm{Z},\beta_m)-\log\pi_m(e)\}}{1+\exp\{\eta_m(e;\bm{Z},\beta_m)-\log\pi_m(e)\}}.
    \end{aligned}
\end{equation*}
Thus, conditioning on inclusion introduces the known offset $-\log\pi_m(e)$ into the original log odds. Similar offsets appear in local case-control sampling \citep{fithian2014local} and noise-contrastive estimation \citep{gutmann2012nce}, whereas the proposal $q_m$ of CINE is built from lower-order scores and requires no evaluation of order-$m$ candidate hyperedges in advance. The offset $-\log\pi_m(e)$ applies to both present hyperedges and sampled negative examples and remains fixed during parameter estimation. Conditional on the preliminary estimate, the Poisson construction yields independent candidate counts, which are also independent of the target-order hyperedge indicators. Combined with the assumed independence of the hyperedge indicators, these properties imply that, conditional on the inclusion indicators, the outcomes of the included candidates are independent Bernoulli variables with the probabilities given above. Taking the negative logarithm of their product likelihood yields the order-$m$ corrected loss, namely the conditional negative log-likelihood:
\begin{equation}\label{eq:corrected-loss}
    \begin{aligned}
        \widehat \ell_m(\bm{Z},\beta_m)=\sum_{e\in E_m^+\cup\widehat E_m^-}&\Bigl[\log\left\{1+\exp\left(\eta_m(e;\bm{Z},\beta_m)-\log\pi_m(e)\right)\right\}\\
        &-A_e^{(m)}\left\{\eta_m(e;\bm{Z},\beta_m)-\log\pi_m(e)\right\}\Bigr].
    \end{aligned}
\end{equation}
For $m=2$, every order-two candidate is included, so we set $\pi_2(e)=1$ and $\widehat\ell_2$ coincides with the order-two loss. Accordingly, we define the CINE estimator by:
\begin{equation}\label{eq:cine}
    \left(\widehat{\bm{Z}},\{\widehat\beta_m\}_{m\in\mathcal{M}}\right)\in\underset{\substack{\bm{Z},\{\beta_m\}_{m\in\mathcal{M}}\\\sum_{i=1}^n \bm{z}_i=\bm{0}}}{\operatorname{arg\,min}}\sum_{m\in\mathcal{M}}\widehat\ell_m(\bm{Z},\beta_m).
\end{equation}
Algorithm~\ref{alg:cine} summarizes the steps of CINE.
\begin{rem}[Sequential refinement: CINE++]\label{rem:cine++}
    CINE constructs the proposal distributions for all higher orders from the same preliminary estimate $\widetilde{\bm{Z}}$. A sequential extension, denoted by CINE++, processes the retained orders in increasing order: for each $m\in\mathcal{M}$ with $m>2$, CINE++ constructs $q_m(e)$ in \eqref{eq:proposal} from the latest latent position estimate based on all lower retained orders, samples negative examples as in CINE, and jointly updates the shared latent positions and order-specific intercepts using all retained orders up to $m$. Information accumulated across orders thus guides subsequent negative sampling, and Section~\ref{Section-Simulation} shows that CINE++ further improves estimation accuracy.
\end{rem}

\subsection{Computation}\label{subsec:computation}
This subsection compares the computational cost of CINE with that of full estimation, starting from a single candidate hyperedge. For $e\in\Omega_m$, the identity $\sum_{\{i,j\}\subset e}\boldsymbol z_i^\top\boldsymbol z_j=\{\|\sum_{i\in e}\boldsymbol z_i\|^2-\sum_{i\in e}\|\boldsymbol z_i\|^2\}/2$ allows the score $s_m(e;\boldsymbol Z)$ to be evaluated in $O(mr)$ operations. The same vector sum can be reused to compute $\partial s_m(e;\boldsymbol Z)/\partial\boldsymbol z_i=(\sum_{j\in e}\boldsymbol z_j-\boldsymbol z_i)/\{r\binom{m}{2}\}^{1/2}$ for $i\in e$, with zero derivatives for nodes outside $e$. Each of these $m$ gradient vectors requires $O(r)$ operations, so evaluating the loss contribution and its gradient for one order-$m$ candidate requires $O(mr)$ operations.

We now analyze the computational complexity of each stage of CINE. First, CINE obtains the preliminary estimate $\widetilde{\boldsymbol Z}$ by minimizing the order-two loss. This step involves all $\binom{n}{2}$ order-two candidates, so each evaluation of the order-two loss and its gradient costs $O(n^2r)$. Second, for each $m\in\mathcal M$ with $m>2$, CINE evaluates $s_{h_m}(u;\widetilde{\boldsymbol Z})$ for every $u\in\Omega_{h_m}$. Each score requires $O(h_mr)$ operations, and exponentiating and normalizing the scores to obtain $q_m^{\mathrm{sub}}(u)$ requires $O\{\binom{n}{h_m}\}$ additional operations. Thus, proposal construction costs $O\{rh_m\binom{n}{h_m}\}$ per target order. Third, generating each order-$m$ candidate and checking whether it is present involves processing its $m$ node indices and requires $O(m)$ expected operations. Processing $N_m$ candidates therefore has an expected sampling cost of $O(mB_m)$ per target order, since $\mathbb E(N_m)=B_m$. Finally, CINE jointly estimates $\widehat{\boldsymbol{Z}}$ and $\{\widehat{\beta}_m\}_{m\in\mathcal{M}}$ using the included candidates $E_m^+\cup\widehat E_m^-$. Since each candidate requires $O(mr)$ operations and $\mathbb E|\widehat E_m^-|\le B_m$, each evaluation of the CINE loss and its gradient has an expected cost of $O[r\sum_{m\in\mathcal M}m(|E_m^+|+B_m)]$. Therefore, counting one evaluation of each loss, the overall expected computational cost of CINE is $O[rn^2+\sum_{m\in\mathcal M} \{rh_m\binom{n}{h_m}+mr(|E_m^+|+B_m)\}]$, accounting for the preliminary estimate, proposal construction, sampling, and final optimization.

For full estimation, each loss and gradient evaluation involves all $\binom{n}{m}$ candidates at every order $m\in\mathcal M$. Applying the same $O(mr)$ computation to each candidate gives a cost of $O\{r\sum_{m\in\mathcal M}m\binom{n}{m}\}$ per evaluation. We next compare the computational costs of the two methods as $n\to\infty$. For a fixed collection of hyperedge orders, $h_m<m$ implies $h_m\binom{n}{h_m}/\{m\binom{n}{m}\}=(h_m/m)\prod_{j=h_m+1}^{m}j/(n-j+1)=O(n^{h_m-m})\to0$. If $\mathcal M$ contains at least one order greater than two, the cost of the preliminary estimate also satisfies $n^2/\{\sum_{m\in\mathcal M}m\binom{n}{m}\}\to0$. Furthermore, if $|E_m^+|+B_m=o\{\binom{n}{m}\}$ for every $m\in\mathcal M$, then $\{\sum_{m\in\mathcal M}m(|E_m^+|+B_m)\}/ \{\sum_{m\in\mathcal M}m\binom{n}{m}\}\to0$. Combining these relations gives $rn^2+\sum_{m\in\mathcal M} \{rh_m\binom{n}{h_m}+mr(|E_m^+|+B_m)\} =o\{r\sum_{m\in\mathcal M}m\binom{n}{m}\}$, which establishes that CINE has a much smaller asymptotic computational cost than full estimation. Here, the condition $|E_m^+|+B_m=o\{\binom{n}{m}\}$ only requires that the number of present hyperedges and the sampling budget grow more slowly than the number of candidate hyperedges, which is a mild requirement in an asymptotically sparse regime.

CINE thus saves computation in two ways: each higher-order loss involves only the present hyperedges and the sampled negative examples, and the proposal is built from lower-order candidates, whose number grows much more slowly than that of the target-order candidates. This combination also sets CINE apart from existing strategies. Truncating and Projecting change the data and hence the estimation target, whereas CINE keeps every present hyperedge and the target of full estimation (Proposition~\ref{prop:2}). Uniform negative sampling keeps the target, but most of its draws are uninformative, and local case-control sampling finds informative negative examples only by scoring every target-order candidate. CINE finds them through lower-order scores, and Section~\ref{Section-Theory} shows that CINE preserves the order of the error bounds of full estimation.

\begin{algorithm}[!htbp]
\small
\setstretch{1.0}
\caption{Cross-order information-guided negative-sampling estimation (CINE)}
\label{alg:cine}
\algrenewcommand{\algorithmicrequire}{\textbf{Input:}}
\algrenewcommand{\algorithmicensure}{\textbf{Output:}}
\algrenewcommand{\algorithmicindent}{1.25em}
\begin{algorithmic}[1]
\setlength{\itemsep}{1pt}
\setlength{\parsep}{0pt}
\setlength{\topsep}{2pt}

\Require Node set $[n]$; retained orders $\mathcal M$ with $2\in\mathcal M$; sets of present hyperedges $\{E_m^+\}_{m\in\mathcal M}$; latent dimension $r$; sampling budgets $\{B_m:m\in\mathcal M\setminus\{2\}\}$; mixture weight $\epsilon$.

\Ensure Estimated latent positions $\widehat{\boldsymbol Z}$ and intercepts $\{\widehat\beta_m\}_{m\in\mathcal M}$.

\State Minimize the order-two loss to obtain $(\widetilde{\boldsymbol Z},\widetilde\beta_2)$.

\For{each $m\in\mathcal M\setminus\{2\}$}

    \State Compute $q_m^{\mathrm{sub}}$ from the scores on $\Omega_{h_m}$ evaluated at the fixed $\widetilde{\boldsymbol Z}$, and obtain $q_m$ from \eqref{eq:proposal}.

    \State Draw $N_m\sim\operatorname{Poisson}(B_m)$ and sample $N_m$ candidates independently with replacement from $q_m$.

    \State Remove duplicate candidates and present hyperedges to obtain $\widehat E_m^-$.

    \State Evaluate $q_m(e)$ and $\pi_m(e)$ for every $e\in E_m^+\cup\widehat E_m^-$.

\EndFor

\State Jointly minimize the order-two loss plus the corrected losses in \eqref{eq:corrected-loss}, obtaining the CINE estimators $\{\widehat{\boldsymbol{Z}},\{\widehat{\beta}_m\}_{m\in\mathcal{M}}\}$.

\end{algorithmic}
\end{algorithm}

In practice, we implement Algorithm~\ref{alg:cine} using L-BFGS-B with analytic gradients and enforce the centering constraint through a reduced-coordinate parameterization. Section~B.1 of the Supplementary Material gives the initialization and stopping criteria.

\begin{rem}[Choosing the retained orders under a time budget]\label{rem:order-selection}
    The cost analysis above gives a simple guide for choosing $\mathcal M$. After computing $\widetilde{\bm Z}$, we record the time $t_{\mathrm{ref}}$ of building the order-three proposal from the $\binom{n}{2}$ order-two candidates. Since building a proposal from order-$h$ candidate hyperedges costs $O\{rh\binom{n}{h}\}$, ignoring constants predicts the time $\widehat t(h)=t_{\mathrm{ref}}\,h\binom{n}{h}/\{2\binom{n}{2}\}$ for a proposal built from order $h$. Since proposal construction is the step whose cost grows fastest with the order, a retained set $\mathcal M$ needs roughly $\sum_{m\in\mathcal M,\,m>2}\widehat t(h_m)$ for proposal construction beyond the time already spent on $\widetilde{\bm Z}$. Given a time budget, we add orders from low to high while this predicted time stays within the budget. Because the proposal of an order can be built from any lower retained order, an intermediate order can be skipped to keep a higher order within the budget. For example, $\mathcal M=\{2,3,4,5,7\}$ in Example 1 with $(n,m_{\max})=(300,7)$ avoids building a proposal from all order-six candidates. Once $\mathcal M$ is fixed, the remaining time determines the budget ratio, because the costs of sampling and loss evaluation grow with $B_m$. With $n=100$ and $m_{\max}=7$, the set $\{2,3,4,5,7\}$ already beats Uniform, Truncating, Projecting, and HGE in latent position recovery within about two seconds (Figure~\ref{fig:retained_orders_matched}).
\end{rem}

%%%%%%%%%%%%%%%%%%%%%%%%%%%%%%%%%%%%%%%%%%%

\section{Theory}
\label{Section-Theory}
In this section, we provide theoretical guarantees for CINE. Throughout this section, the retained order set $\mathcal M$ is fixed and does not depend on $n$. We first introduce some notation. Let $\bm Z^*=(\bm z_1^*,\ldots,\bm z_n^*)^\top$ and $\{\beta_m^*\}_{m\in\mathcal M}$ denote the true latent position matrix and order-specific intercepts. Given $\bm Z^*$, the conditional expectation of the number of order-$m$ present hyperedges is $\mu_m=\mathbb{E}(|E_m^+|\mid\bm{Z}^*)=\sum_{e\in\Omega_m}p_m(e;\bm{Z}^*,\beta_m^*)$. We further define $\varrho_{m,n}= \mu_m/\binom nm = \sum_{e\in\Omega_m}p_m(e;\bm Z^*,\beta_m^*)/\binom nm$, which is the expected proportion of present hyperedges among these candidates. For each $m\in\mathcal M$ with $m>2$, let $F_m$ denote the full-data information, namely the Fisher information of the order-$m$ component of the full Bernoulli loss in \eqref{eq:full-loss}, and let $J_m$ denote the expected information contributed by the corrected loss in \eqref{eq:corrected-loss}. These matrices are given by:
\begin{equation*}
    \begin{aligned}
        F_m&=\sum_{e\in\Omega_m}p_m(e;\bm{Z}^*,\beta_m^*)\{1-p_m(e;\bm{Z}^*,\beta_m^*)\}\nabla\eta_m(e;\bm{Z}^*,\beta_m^*)\nabla\eta_m(e;\bm{Z}^*,\beta_m^*)^{\top},\\
        J_m&=\sum_{e\in\Omega_m}\frac{p_m(e;\bm{Z}^*,\beta_m^*)\{1-p_m(e;\bm{Z}^*,\beta_m^*)\}\pi_m(e)}{p_m(e;\bm{Z}^*,\beta_m^*)+\{1-p_m(e;\bm{Z}^*,\beta_m^*)\}\pi_m(e)}
    \nabla\eta_m(e;\bm{Z}^*,\beta_m^*)\nabla\eta_m(e;\bm{Z}^*,\beta_m^*)^{\top},
    \end{aligned}
\end{equation*}
where $\nabla\eta_m(e;\bm Z^*,\beta_m^*)$ denotes the gradient of the log odds $\eta_m(e;\bm Z,\beta_m)$ with respect to the latent-position and intercept parameters, evaluated at $(\bm Z^*,\beta_m^*)$. The matrix $F_m$ quantifies the information available from all order-$m$ candidates, whereas $J_m$ measures the information retained by the corrected loss after CINE sampling.

\subsection{Sampling Properties}\label{subsec:sampling}
We first study the theoretical properties of the proposal distribution. We begin with the following assumption on the preliminary estimate.

\begin{assumption}[Independence of the proposal]\label{ass:1}
For every $m\in\mathcal M$ with $m>2$, the latent position estimate $\widetilde{\boldsymbol{Z}}$ used to construct $q_m$ is independent of the order-$m$ hyperedge indicators \(\{A_e^{(m)}:e\in\Omega_m\}\) and satisfies $\mathbb{P}\{\|\widetilde{\boldsymbol{Z}}\|_{\mathrm F}<\infty\}=1$.
\end{assumption}

Assumption \ref{ass:1} has two components. First, the independence condition ensures that, conditional on $\widetilde{\boldsymbol{Z}}$, the order-$m$ indicators remain mutually independent and retain their original occurrence probabilities $p_m(e;\boldsymbol{Z}^*,\beta_m^*)$. CINE and CINE++ both satisfy this condition, because the estimate for order $m$ uses only lower orders, so all results in this section apply to both methods, with $\widetilde{\boldsymbol{Z}}$ denoting the estimate that constructs $q_m$. Second, the finiteness condition makes every $\exp\{\tau_m s_{h_m}(u;\widetilde{\boldsymbol{Z}})\}$ and their sum over the finite set $\Omega_{h_m}$ finite and strictly positive almost surely, so the probabilities $q_m^{\mathrm{sub}}(u)$ are valid. Based on this assumption, we establish the following properties of the proposed sampling strategy.

\begin{proposition}[Preservation of the population target]\label{prop:2}
Under Assumption~\ref{ass:1}, for every \(m\in\mathcal M\) with \(m>2\) and every finite alternative \((\bm Z',\beta_m')\), the following inequality holds almost surely:
\begin{equation*}\label{eq:population-target}
    \mathbb{E}\{\widehat\ell_m(\bm Z',\beta_m') -\widehat\ell_m(\bm Z^*,\beta_m^*)\mid \widetilde{\boldsymbol{Z} }\}\ge0.
\end{equation*}
The equality holds if and only if $ p_m(e;\bm Z',\beta_m')=p_m(e;\bm Z^*,\beta_m^*)$ for every $e\in\Omega_m.$ Moreover, we have:
 \begin{equation*}
     \begin{aligned}
 \bigl(\bm Z^*,\{\beta_m^*\}_{m\in\mathcal M}\bigr)=\operatorname*{arg\,min}_{\substack{\bm Z',\,\{\beta_m'\}_{m\in\mathcal M}\\\sum_{i=1}^{n}\bm z_i'=0}}
 \mathbb{E}\!\left\{
   \sum_{m\in\mathcal M}\widehat\ell_m(\bm Z',\beta_m')
 \right\}=
 \operatorname*{arg\,min}_{\substack{\bm Z',\,\{\beta_m'\}_{m\in\mathcal M}\\\sum_{i=1}^{n}\bm z_i'=0}}
 \mathbb{E}\!\left\{
   \ell\!\left(\bm Z',\{\beta_m'\}_{m\in\mathcal M}\right)
 \right\}.
\end{aligned}
 \end{equation*}
\end{proposition}

Proposition~\ref{prop:2} shows that the expected CINE loss and the expected full Bernoulli loss have the same minimizers. Thus, although our sampling strategy increases the proportion of present hyperedges in the estimation sample, the offset $-\log\pi_m(e)$ in \eqref{eq:corrected-loss} preserves the optimization objective of the full Bernoulli loss. Therefore, Proposition~\ref{prop:2} provides a theoretical justification for estimating $\bm Z$ and $\{\beta_m\}_{m\in\mathcal M}$ by minimizing the CINE loss constructed from \eqref{eq:corrected-loss}. We next quantify how much of the full-data information CINE retains.

\begin{theorem}[Information guarantee]
\label{thm:information}
Under Assumption~\ref{ass:1}, for every $m\in\mathcal{M}$ with \(m>2\) and every fixed sampling budget $0 < B_m<+\infty$, if there exists a constant $\rho_m>0$ such that $\pi_m(e)\ge \rho_m p_m(e;\boldsymbol Z^*,\beta_m^*)$ for every $e\in\Omega_m$, then we have:
\begin{equation*}
\label{eq:information-coverage}
    \frac{\rho_m}{1+\rho_m}F_m \preceq J_m\preceq F_m.
\end{equation*}
\end{theorem}

Theorem~\ref{thm:information} relates the information retained by CINE to that available from all order-$m$ candidate hyperedges. The order-specific quantity $\rho_m$ controls a lower bound on the fraction of information retained. When $\rho_m$ is large, the bound guarantees that negative sampling incurs only a small relative loss of information. Since the order-two loss enters in full, every retained higher order adds at least a fraction $\rho_m/(1+\rho_m)$ of its full-data information about the shared latent positions. Corollary~\ref{cor:information-budget} expresses $\rho_m$ through the sampling budget.

\begin{corollary}
\label{cor:information-budget}
Under the conditions of Theorem~\ref{thm:information}, if there exists a constant \(a\in(0,1]\) such that $q_m(e)\ge a\,p_m(e;\boldsymbol Z^*,\beta_m^*)/\mu_m$ for every $e\in\Omega_m$, then almost surely, the conclusion in Theorem~\ref{thm:information} holds for a choice of \(\rho_m\) satisfying $\rho_m\ge \frac{aB_m}{\mu_m+aB_mp_{m,\max}}$, where $p_{m,\max}:=\max_{e\in\Omega_m}p_m(e;\boldsymbol Z^*,\beta_m^*)$. Moreover, we have:
\begin{equation*}
    \frac{aB_m}{\mu_m+aB_m}F_m\preceq J_m\preceq F_m.
\end{equation*}
\end{corollary}

The additional condition in Corollary~\ref{cor:information-budget}, $q_m(e)\ge a\,p_m(e;\boldsymbol Z^*,\beta_m^*)/\mu_m$, requires the proposal probability assigned to each candidate hyperedge to be at least a proportion $a$ of its normalized occurrence probability. This condition holds automatically: Proposition~A.2 in the Supplementary Material shows that it holds for every $\widetilde{\boldsymbol{Z}}$ with $a$ bounded away from zero when the true scores are bounded and $\epsilon>0$, and that the second term of \eqref{eq:proposal} increases $a$ as $\widetilde{\boldsymbol{Z}}$ becomes more accurate and as lower-order scores track higher-order scores. Under this condition, Corollary \ref{cor:information-budget} bounds $\rho_m$ from below in terms of $B_m$, $\mu_m$, and $p_{m,\max}$. In particular, for fixed $a>0$, the guaranteed fraction of full-data information retained approaches one as $B_m/\mu_m\to\infty$. For example, if the proposal probabilities match the normalized occurrence probabilities ($a=1$), the sampling budget $B_m=4\mu_m$ used in Section~\ref{Section-Simulation} retains at least 80\% of the full-data information at each target order. The budget ratio thus trades computation time for retained information, as the term $n/B_m$ in the error levels of Theorem~\ref{thm:accuracy} below also shows for the estimation error.

\subsection{Nonasymptotic Error Bounds}
In this subsection, we establish nonasymptotic error bounds for the CINE estimators $\widehat{\boldsymbol Z}$ and $\{\widehat{\beta}_m\}_{m\in\mathcal M}$. Under the model, the true parameters $\boldsymbol Z^*$ and $\{\beta_m^*\}_{m\in\mathcal M}$ minimize the expected full Bernoulli loss $\mathbb E\{\ell(\boldsymbol Z,\{\beta_m\}_{m\in\mathcal M})\}$ subject to the centering constraint $\boldsymbol1^\top\boldsymbol Z=0$. The analysis relies on the following assumptions.

\begin{assumption}
\label{ass:2}
For fixed constants \(L>0\) and \(c_p\in(0,1)\), $\max_{i\in[n]}\|\boldsymbol z_i^*\|_2\le L$ and $\max_{m\in\mathcal M}\max_{e\in\Omega_m} p_m(e;\boldsymbol Z^*,\beta_m^*)\le1-c_p$.
\end{assumption}
The first condition in Assumption~\ref{ass:2} bounds the $\ell_2$-norms of the true latent positions, ensuring that the scores $s_m(e;\boldsymbol Z^*)$ remain uniformly bounded as the number of nodes increases. Related bounds on latent factors are used by \citet{yuan2023high}. The second condition requires hyperedge occurrence probabilities to be uniformly bounded away from one. Consequently, $0<\varrho_{m,n}\le1-c_p$, so the expected proportion of present hyperedges among order-$m$ candidate hyperedges is also bounded away from one. This condition holds naturally for the sparse hypergraphs considered here. A related upper bound on edge probabilities is imposed by \citet{ma2020universal} in their analysis of latent space models.

\begin{assumption}\label{ass:3}
    There exists a positive constant $c_1$ such that $\lambda_{\min}\{(\boldsymbol Z^*)^\top \boldsymbol Z^*/n\}\ge c_1$.
\end{assumption}

Assumption~\ref{ass:3} imposes a lower bound on the eigenvalues of $(\boldsymbol Z^*)^\top\boldsymbol Z^*/n$. This condition ensures that $\boldsymbol Z^*$ has full column rank, allowing stable recovery of the latent positions from their inner products up to an orthogonal transformation. A similar eigenvalue condition appears in \citet{Wu27082026}. The following theorem establishes an upper bound on the estimation error of the CINE estimators.

\begin{theorem}[Nonasymptotic error bounds]\label{thm:accuracy}
Under Assumptions~\ref{ass:2} and~\ref{ass:3}, if there exists a sufficiently large constant \(K>0\) such that $n\varrho_{2,n}\ge K\log n$, $n^{m-1}\varrho_{m,n}\ge K\log n$ and $B_m\ge Kn\log n$ hold for $m\in\mathcal{M}\setminus\{2\}$, then, with the error levels $\omega_{2,n}=1/(n\varrho_{2,n})$ and $\omega_{m,n}=1/(n^{m-1}\varrho_{m,n})+n/B_m$ for $m>2$, the following bounds hold with probability at least \(1-C/n\) for a constant \(C>0\) independent of \(n\):
\begin{subequations}\label{eq:direct-parameter-bounds}
\begin{align}
 \frac1{\sqrt n}\inf_{O^\trans O=I_r}\|\widehat{\boldsymbol Z}O-\boldsymbol Z^*\|_{\mathrm F}&\le C\sqrt{\log^3 n\,\min_{m\in\mathcal M}\omega_{m,n}},\label{eq:position-root-bound}\\
 |\widehat\beta_m-\beta_m^*|&\le C\sqrt{\log n\,\omega_{m,n}},\qquad m\in\mathcal M.\label{eq:intercept-bound}
\end{align}
\end{subequations}
\end{theorem}

Theorem~\ref{thm:accuracy} provides upper bounds on the estimation errors of the CINE estimators, and these bounds hold for every centered parameter whose CINE loss is at most that at the truth, not only for the global minimizer. Each retained order $m$ has its own error level $\omega_{m,n}$, and the latent positions are estimated at the level of the most informative retained order, so retaining a higher order can only tighten the bound in \eqref{eq:position-root-bound}. Each intercept $\beta_m$ is estimated at the level of its own order. For $m>2$, the first term of $\omega_{m,n}$ reflects the limited number of order-$m$ present hyperedges, and the second, $n/B_m$, comes from negative sampling, which reduces the effective number of order-$m$ observations from $\mu_m$ to $\mu_mB_m/(\mu_m+B_m)$. When $\log n\,\omega_{m,n}=o(1)$ for every $m\in\mathcal M$ and $\log^3 n\,\min_{m\in\mathcal M}\omega_{m,n}=o(1)$, all bounds in Theorem~\ref{thm:accuracy} vanish as $n\to\infty$, establishing the consistency of the CINE estimator up to an orthogonal transformation of the latent positions. Section~A.6 of the Supplementary Material shows that these bounds hold simultaneously when the retained orders are selected from the data among finitely many orders, as in Section~\ref{Section-Real-Data}.

To assess the effect of negative sampling on parameter estimation, we compare the error bound of CINE estimators with that of full estimation. Specifically, we define the full estimator $\{\bar{\boldsymbol Z},\{\bar\beta_m\}_{m\in\mathcal M}\}$ of the latent position matrix and the order-specific intercepts as a minimizer of the full Bernoulli loss $\ell(\boldsymbol Z,\{\beta_m\}_{m\in\mathcal M})$ in \eqref{eq:full-loss} subject to $\sum_{i=1}^n\boldsymbol z_i=\boldsymbol 0$. The following proposition bounds the estimation error of the full estimator.

\begin{proposition}[Error bounds for the full estimator]\label{lem:full-data}
Under Assumptions~\ref{ass:2} and~\ref{ass:3}, suppose \(n^{m-1}\varrho_{m,n}\ge K\log n\) for every \(m\in\mathcal M\), including \(m=2\), with \(K>0\) sufficiently large. With probability at least \(1-C/n\) over the hyperedge indicators conditional on \(\boldsymbol Z^*\), for a constant \(C>0\) independent of \(n\), the full estimator satisfies the bounds in Theorem~\ref{thm:accuracy} with the term $n/B_m$ removed from every $\omega_{m,n}$.
\end{proposition}
Theorem~\ref{thm:accuracy} and Proposition~\ref{lem:full-data} differ only in the term $n/B_m$ of $\omega_{m,n}$. If $\mu_m=O(B_m)$, then $n/B_m=O\{1/(n^{m-1}\varrho_{m,n})\}$, so all error bounds of CINE and full estimation have the same order. In particular, when \(B_m\asymp\mu_m\) for each \(m>2\), the computation of the CINE loss uses at most \(\mu_m+B_m=O(\mu_m)\) candidate hyperedges on average, compared with all \(\binom{n}{m}\) candidates of that order for the full Bernoulli loss. CINE therefore preserves the order of the error bounds of full estimation at a much lower computational cost.

\section{Simulation}
\label{Section-Simulation}

We evaluated CINE in three examples, which focus on estimation accuracy and computation time, the effects of sampling budget and signal strength, and the roles of order-two and higher-order information.

\noindent\textbf{Example 1.} We examined how estimation accuracy and computation time varied with $n$ and the maximum hyperedge order $m_{\max}$. We set $n\in\{100,300\}$ and $m_{\max}\in\{3,5,7\}$, with $c=1.8$. Here, $c$ is the signal strength, which controls the scale of $\bm{Z}$. We fixed the expected average participation $m\mu_m/n$, the expected number of order-$m$ hyperedges containing a node, at $2\mu_2/n=12$ for $m=2$ and $m\mu_m/n=20$ for $m=3,\ldots,m_{\max}$. The budget ratio was fixed at $B_m/\mu_m=4$ for $m=3,\ldots,7$.

\noindent\textbf{Example 2.} We examined the effects of sampling budget and signal strength on estimation accuracy. We fixed $n=100$ and $m_{\max}=7$ and varied $c\in\{1.4,1.6,1.8,2.0\}$ and the budget ratio $B_m/\mu_m\in\{4,6,8\}$ for $m=3,\ldots,7$. We fixed the expected average participation at $2\mu_2/n=12$ for $m=2$ and $m\mu_m/n=20$ for $m=3,\ldots,7$.

\noindent\textbf{Example 3.} We examined how estimation accuracy varied with the amount of order-two and higher-order information. We fixed $n=100$, $m_{\max}=7$, and $c=1.8$. We considered all combinations of $2\mu_2/n\in\{8,12,16\}$ and $m\mu_m/n\in\{16,20,24\}$, for $m=3,\ldots,7$. The budget ratio was fixed at $B_m/\mu_m=4$ for $m=3,\ldots,7$.

For each setting, we generated 100 replicates with latent dimension $r=2$. For each node $i$, we set $\bm{z}_i^{\mathrm{raw}} = c\exp(0.12\xi_i)(\cos\theta_i,\sin\theta_i)^\top + \bm{\varepsilon}_i$, where $\theta_i \sim \mathrm{Uniform}(0,2\pi)$, $\xi_i \sim N(0,1)$, and $\bm{\varepsilon}_i \sim N(\bm{0},0.12^2 \bm{I}_2)$ were independent across nodes, and centered the latent positions by $\bm{z}_i = \bm{z}_i^{\mathrm{raw}} - n^{-1}\sum_{j=1}^{n}\bm{z}_j^{\mathrm{raw}}$. For each $m=2,\ldots,m_{\max}$ and $e\in\Omega_m$, we generated $A_e^{(m)} \mid \bm{Z} \sim \operatorname{Bernoulli}\{p_m(e)\}$ independently, with $\beta_m$ chosen such that $\sum_{e\in\Omega_m}p_m(e;\bm{Z},\beta_m)=\mu_m$ for the expected average participation $m\mu_m/n$ specified in each example.

We implemented CINE with $\mathcal M=\{2,\ldots,m_{\max}\}$ as the default set of retained orders. For $(n,m_{\max})=(300,7)$ in Example 1, we used $\mathcal M=\{2,3,4,5,7\}$, chosen as in Remark~\ref{rem:order-selection}. We fixed the mixture weight at $\epsilon=0.20$.

For comparison, we considered the following methods. (1) FULL: Full estimation using all candidates at orders $2,\ldots,m_{\max}$. (2) Uniform: Fixing $\epsilon=1$ in CINE, while keeping the retained orders and sampling budgets unchanged. (3) Truncating: Retaining only the order-two and order-three data and fitting the Bernoulli model using all candidates at these two orders. (4) Projecting: Retaining the original order-two data and projecting each present hyperedge of order at least three onto all its three-node subsets. Repeated triples are treated as a single present hyperedge, and the Bernoulli model is fitted to the resulting order-two and order-three data. (5) Joint link embedding (JLE) in \citet{yuan2023high}. (6) Hypergraph embedding model (HEM) in \citet{zhen2023community}. (7) The latent embedding approach of \citet{Wu27082026}, hereafter referred to as HGE. For HGE, we keep its estimated node embeddings fixed and obtain occurrence probabilities from \eqref{eq:model}, as in CINE.

We use three performance measures. (1) Latent position recovery: Let $\widehat{\bm{Z}}$ denote the estimated latent position matrix after centering, and let $\bm{O}^*=\arg\min_{\bm{O}^\top \bm{O}=\bm{I}_r}\|\widehat{\bm{Z}}\bm{O}-\bm{Z}\|_F^2$. We calculate the relative mean squared error (relative MSE) $\mathrm{RelMSE}_{\bm{Z}}=\|\widehat{\bm{Z}}\bm{O}^*-\bm{Z}\|_F^2/\|\bm{Z}\|_F^2.$ (2) Occurrence probability recovery: Let $\widehat p_m(e)$ denote the estimated occurrence probability of candidate $e$ of order $m$. For HEM and JLE, we use the probabilities defined by their respective fitted models. For JLE, we include a separate intercept for each fitted order to accommodate differences in sparsity across orders. For the remaining methods, $\widehat p_m(e)=p_m(e;\widehat{\bm{Z}},\widehat\beta_m)$ with each method's estimates. For each evaluated order, we calculate $\mathrm{RelMSE}_{p,m}=\{\sum_{e\in\Omega_m}[\widehat p_m(e)-p_m(e)]^2\}/\{\sum_{e\in\Omega_m}p_m(e)^2\}.$ We also report the overall relative mean squared error across all orders, $\mathrm{RelMSE}_{p,\mathrm{all}}=\{\sum_{m=2}^{m_{\max}}\sum_{e\in\Omega_m}[\widehat p_m(e)-p_m(e)]^2/ \{\sum_{m=2}^{m_{\max}}\sum_{e\in\Omega_m}p_m(e)^2\}.$ (3) Computation time: We record the wall-clock time of each method, including initialization. To reduce the computational cost arising from large candidate sets, we used accelerated procedures for data generation and probability evaluation, described in Section~B.2 of the Supplementary Material.

\begin{table}[!t]
    \centering
    \singlespacing
    \caption{Relative mean squared errors and computation time in Example 1 with maximum hyperedge order $m_{\max}=5$. Entries are means (standard deviations) over 100 replicates. Among methods other than FULL, the lowest mean relative MSE in each column of each panel is shown in bold and the second lowest is underlined.}
    \label{tab:example1_m5}
    \begingroup
    \fontsize{8}{9.6}\selectfont
    \renewcommand{\arraystretch}{1.00}
    \setlength{\tabcolsep}{1.5pt}
    \setlength{\aboverulesep}{0.35ex}
    \setlength{\belowrulesep}{0.35ex}
    \begin{tabular*}{\linewidth}{@{\extracolsep{\fill}}lccccccc@{}}
        \toprule
        \multicolumn{8}{@{}l}{Panel A: $n=100$.} \\
        \midrule
        \multirow{2}{*}{\textsc{Method}} & \multicolumn{6}{c}{Relative MSE} & \multirow{2}{*}{\textsc{Time} (s)} \\
        \cmidrule(lr){2-7}
        & $\bm{Z}$ & Overall & Order~2 & Order~3 & Order~4 & Order~5 & \\
        \midrule
        \textit{FULL} & \textit{0.013}\,\textit{(0.002)} & \textit{0.032}\,\textit{(0.003)} & \textit{0.025}\,\textit{(0.003)} & \textit{0.065}\,\textit{(0.012)} & \textit{0.093}\,\textit{(0.024)} & \textit{0.115}\,\textit{(0.038)} & \textit{187.581}\,\textit{(57.363)} \\
        \textsc{Cine++} & $\mathbf{0.020}$\,(0.003) & $\mathbf{0.050}$\,(0.006) & $\mathbf{0.039}$\,(0.004) & $\mathbf{0.111}$\,(0.026) & $\mathbf{0.165}$\,(0.051) & $\mathbf{0.200}$\,(0.068) & $0.416$\,(0.033) \\[0.35ex]
        \textsc{Cine} & $\underline{0.022}$\,(0.004) & $\underline{0.059}$\,(0.010) & $\underline{0.045}$\,(0.006) & $\underline{0.133}$\,(0.038) & $\underline{0.210}$\,(0.102) & $\underline{0.264}$\,(0.155) & $0.780$\,(0.269) \\[0.35ex]
        \textsc{Uniform} & $0.029$\,(0.004) & $0.129$\,(0.040) & $0.065$\,(0.008) & $0.380$\,(0.121) & $1.129$\,(0.706) & $2.224$\,(1.947) & $0.545$\,(0.190) \\[0.35ex]
        \textsc{Truncating} & $0.030$\,(0.004) & $0.076$\,(0.008) & $0.057$\,(0.006) & $0.170$\,(0.036) & $0.273$\,(0.074) & $0.364$\,(0.125) & $0.472$\,(0.137) \\[0.35ex]
        \textsc{Projecting} & $0.023$\,(0.004) & $0.116$\,(0.073) & $0.047$\,(0.007) & $0.320$\,(0.150) & $1.236$\,(0.978) & $2.845$\,(3.034) & $0.459$\,(0.133) \\[0.35ex]
        \textsc{JLE} & $0.442$\,(0.155) & $1.348$\,(2.127) & $0.569$\,(0.434) & $2.853$\,(4.922) & $18.901$\,(49.314) & $128.818$\,(364.153) & $1493.131$\,(361.728) \\[0.35ex]
        \textsc{HEM} & $0.621$\,(0.215) & $1.000$\,(0.000) & $1.000$\,(0.000) & $0.999$\,(0.000) & $0.991$\,(0.003) & $0.990$\,(0.014) & $801.447$\,(143.901) \\[0.35ex]
        \textsc{HGE} & $0.058$\,(0.025) & $0.298$\,(0.123) & $0.121$\,(0.025) & $0.937$\,(0.365) & $3.582$\,(2.880) & $8.700$\,(10.412) & $6.658$\,(2.732) \\[0.35ex]
        \midrule
        \multicolumn{8}{@{}l}{Panel B: $n=300$.} \\
        \midrule
        \multirow{2}{*}{\textsc{Method}} & \multicolumn{6}{c}{Relative MSE} & \multirow{2}{*}{\textsc{Time} (s)} \\
        \cmidrule(lr){2-7}
        & $\bm{Z}$ & Overall & Order~2 & Order~3 & Order~4 & Order~5 & \\
        \midrule
        \textit{FULL}\textsuperscript{\dag} & - & - & - & - & - & - & - \\
        \textsc{Cine++} & $\mathbf{0.016}$\,(0.001) & $\mathbf{0.054}$\,(0.004) & $\underline{0.051}$\,(0.004) & $\mathbf{0.117}$\,(0.014) & $\mathbf{0.170}$\,(0.027) & $\mathbf{0.213}$\,(0.041) & $13.091$\,(5.255) \\[0.35ex]
        \textsc{Cine} & $\underline{0.017}$\,(0.001) & $\underline{0.058}$\,(0.004) & $0.054$\,(0.004) & $\underline{0.127}$\,(0.015) & $\underline{0.184}$\,(0.030) & $\underline{0.234}$\,(0.041) & $11.292$\,(5.201) \\[0.35ex]
        \textsc{Uniform} & $0.025$\,(0.002) & $0.114$\,(0.013) & $0.099$\,(0.009) & $0.415$\,(0.117) & $0.889$\,(0.357) & $1.462$\,(0.718) & $2.904$\,(0.357) \\[0.35ex]
        \textsc{Truncating} & $0.027$\,(0.002) & $0.085$\,(0.007) & $0.080$\,(0.006) & $0.191$\,(0.023) & $0.282$\,(0.045) & $0.373$\,(0.071) & $8.835$\,(1.735) \\[0.35ex]
        \textsc{Projecting} & $0.018$\,(0.001) & $\mathbf{0.054}$\,(0.006) & $\mathbf{0.050}$\,(0.004) & $\underline{0.127}$\,(0.039) & $0.205$\,(0.118) & $0.289$\,(0.233) & $6.332$\,(1.151) \\[0.35ex]
        \textsc{JLE}\textsuperscript{\dag} & - & - & - & - & - & - & - \\[0.35ex]
        \textsc{HEM}\textsuperscript{\dag} & - & - & - & - & - & - & - \\[0.35ex]
        \textsc{HGE} & $0.042$\,(0.023) & $0.195$\,(0.088) & $0.158$\,(0.050) & $0.881$\,(0.750) & $2.768$\,(4.051) & $5.559$\,(9.986) & $65.894$\,(27.516) \\[0.35ex]
        \bottomrule
    \end{tabular*}
    \endgroup
    \par\vspace{0.4em}
    \begin{minipage}{\linewidth}
        \footnotesize\singlespacing
        \textsuperscript{\dag} Panel B omits FULL, JLE, and HEM because each would take over 30 days for 100 replicates.
    \end{minipage}
\end{table}

For Example 1, Table~\ref{tab:example1_m5} shows that CINE offers a good trade-off between estimation accuracy and computation time. In the upper panel ($n=100$ and $m_{\max}=5$), FULL achieved the lowest relative MSEs ($\mathrm{RelMSE}_{\bm{Z}}=0.013, \mathrm{RelMSE}_{p,\mathrm{all}}=0.032$) by using all candidates, but required $187.581$ seconds on average. In contrast, Uniform, Truncating, and Projecting were much faster ($0.459$--$0.545$ seconds) but yielded larger relative MSEs ($\mathrm{RelMSE}_{\bm{Z}}$: $0.023$--$0.030$; $\mathrm{RelMSE}_{p,\mathrm{all}}$: $0.076$--$0.129$). JLE, HEM, and HGE required longer computation times ($6.658$--$1493.131$ seconds) than these three methods and still had larger relative MSEs than CINE. CINE attained lower relative MSEs than all competing methods except FULL ($\mathrm{RelMSE}_{\bm{Z}}=0.022$; $\mathrm{RelMSE}_{p,\mathrm{all}}=0.059$) with a computation time of only $0.780$ seconds, about 240 times faster than FULL. CINE++ further reduced relative MSEs. In the lower panel ($n=300$), latent position recovery improved for all methods reported in both panels, and CINE and CINE++ again attained the lowest latent position errors. Tables~B.1--B.3 in the Supplementary Material show the same pattern.

\begin{figure}[!t]
    \centering
    \begin{minipage}[t]{0.48\linewidth}
        \centering
        \includegraphics[width=\linewidth]{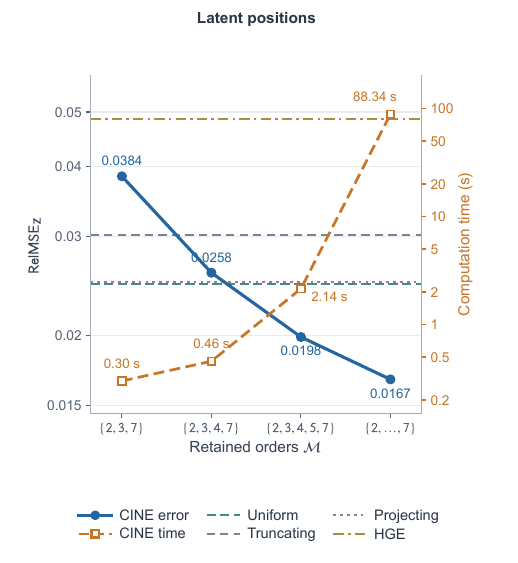}
        \caption{Latent position recovery and computation time of CINE under different retained orders in Example 1 with $n=100$ and $m_{\max}=7$. The blue solid and orange dashed curves show $\mathrm{RelMSE}_{\bm{Z}}$ and computation time, respectively, on logarithmic axes. Horizontal lines mark $\mathrm{RelMSE}_{\bm{Z}}$ of Uniform, Truncating, Projecting, and HGE, whose computation times are $0.57$, $0.40$, $0.43$, and $8.30$ seconds.}
        \label{fig:retained_orders_matched}
    \end{minipage}
    \hfill
    \begin{minipage}[t]{0.48\linewidth}
        \centering
        \includegraphics[width=\linewidth]{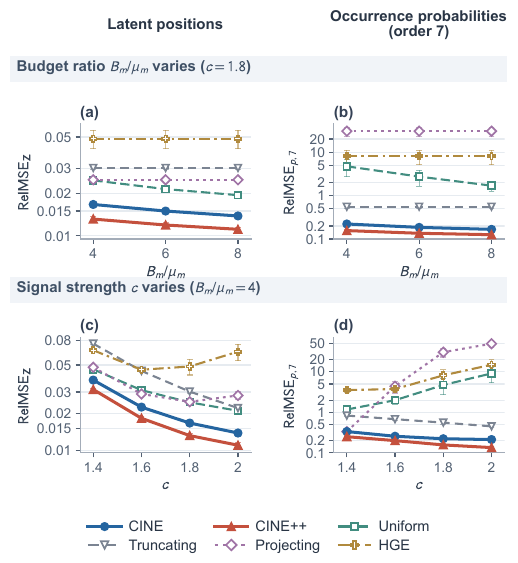}
        \caption{Latent position and order-seven occurrence probability recovery in Example 2. The left and right columns report $\mathrm{RelMSE}_{\bm{Z}}$ and $\mathrm{RelMSE}_{p,7}$, respectively. Panels (a)--(b) vary the budget ratio $B_m/\mu_m$ for $m=3,\ldots,7$ at $c=1.8$, and panels (c)--(d) vary the signal strength $c$ at $B_m/\mu_m=4$.}
        \label{fig:example2_trends}
    \end{minipage}
\end{figure}

Figure~\ref{fig:retained_orders_matched} traces the trade-off of Figure~\hyperref[fig:motivation]{1(e)} along the retained orders of Remark~\ref{rem:order-selection}. We revisit Example 1 with $n=100$ and $m_{\max}=7$ and fit CINE with four nested sets of retained orders, from $\{2,3,7\}$ to $\{2,\ldots,7\}$. Each added order lowers $\mathrm{RelMSE}_{\bm Z}$ but lengthens the computation. With $\mathcal M=\{2,3,4,5,7\}$, CINE attains $\mathrm{RelMSE}_{\bm Z}=0.020$ in $2.14$ seconds, below Uniform, Truncating, and Projecting. CINE is then both faster and more accurate than HGE, which attains $\mathrm{RelMSE}_{\bm Z}=0.049$ in $8.30$ seconds (Table~B.3 in the Supplementary Material). Retaining all orders lowers $\mathrm{RelMSE}_{\bm Z}$ to $0.017$ in $88.34$ seconds.

For Example 2, Figure~\ref{fig:example2_trends} shows how the relative MSEs for latent positions and order-seven occurrence probabilities vary with the budget ratio and the signal strength. Tables~B.4--B.7 in the Supplementary Material report the complete results. At $c=1.8$, raising the budget ratio from four to eight lowered $\mathrm{RelMSE}_{\bm Z}$ of CINE from $0.017$ to $0.014$ and increased its computation time from $88.3$ to $140.5$ seconds (Figure~\hyperref[fig:example2_trends]{4(a)} and Table~B.6), which is the trade-off controlled by the budget ratio. With the budget ratio fixed at four, both relative MSEs decreased for CINE and CINE++ as the signal became stronger. In contrast, HGE, Projecting, and Uniform yielded larger relative MSEs for occurrence probabilities as signal strength increased. For Projecting and HGE, the latent position relative MSEs also increased when $c$ rose from $1.8$ to $2.0$.

\begin{figure}[!h]
    \centering
    \includegraphics[width=\linewidth]{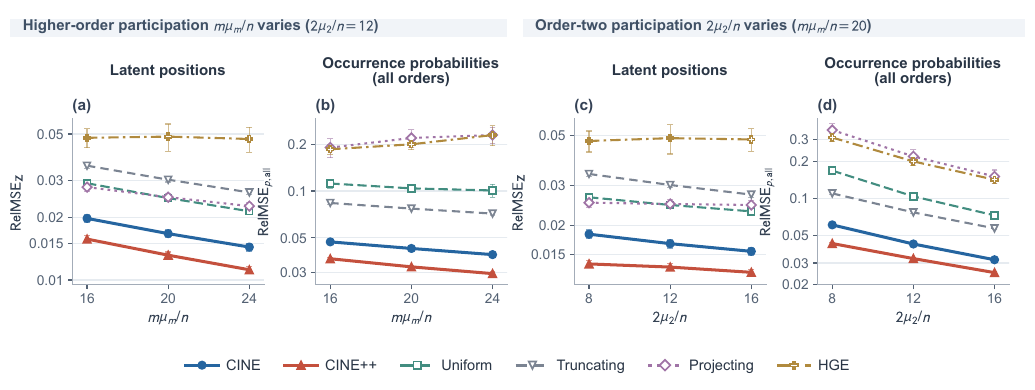}
    \caption{Latent position and overall occurrence probability recovery in Example 3. The left and right columns report $\mathrm{RelMSE}_{\bm{Z}}$ and $\mathrm{RelMSE}_{p,\mathrm{all}}$, respectively. Panels (a)--(b) vary the common higher-order participation $m\mu_m/n$ with order-two participation fixed at $2\mu_2/n=12$. Panels (c)--(d) vary order-two participation with higher-order participation fixed at $m\mu_m/n=20$ for $m=3,\ldots,7$.}
    \label{fig:example3_trends}
\end{figure}

For Example 3, Figure~\ref{fig:example3_trends} summarizes latent position and overall occurrence probability recovery, with higher-order participation varying in panels (a)--(b) and order-two participation varying in panels (c)--(d). Tables~B.8--B.10 in the Supplementary Material report the complete results. Increasing either order-two or higher-order participation with the other held fixed reduced both relative MSEs for CINE and CINE++. For example, with higher-order participation fixed at $m\mu_m/n=20$, increasing order-two participation from $8$ to $16$ reduced CINE's latent position relative MSE from $0.018$ to $0.015$ and its overall probability relative MSE from $0.061$ to $0.032$.

Section~B.4 of the Supplementary Material assesses robustness to model misspecification through pairwise comparisons of CINE with JLE, HGE, and HEM, generating data both from the model of Example 1 and from a mechanism associated with each competing method. Under the competing mechanisms, $\mathrm{RelMSE}_{p,\mathrm{all}}$ of CINE grows by factors of only $3.9$, $1.2$, and $2.1$ in the comparisons with JLE, HGE, and HEM, respectively. Under the model of Example 1, the errors of JLE, HGE, and HEM grow by factors of $72.4$, $1.6$, and $59.4$.
\section{Real Data}\label{Section-Real-Data}
We apply CINE to a coauthorship hypergraph constructed from the Multi-Attribute Dataset on Statisticians (MADStat), which contains bibliographic and citation records for 83,331 papers by 47,311 authors in 36 journals during 1975--2015 \citep{ji2022co}. We focus on papers published in the \emph{Annals of Statistics}, \emph{Biometrika}, \emph{Journal of the American Statistical Association}, and \emph{Journal of the Royal Statistical Society, Series B}, during 2005--2015. Within this collection, we select authors using a coauthorship graph that links pairs with at least two joint papers, regardless of team size. We refer to the 335 authors in its largest connected component as core members. Including all authors of papers involving at least two core members expands this set to 644 authors. We then retain all coauthored papers whose complete author teams lie within this set, yielding 811 papers. These papers define 625 hyperedges, with each distinct complete author team represented by a single hyperedge.

\begin{figure}[htbp]
    \centering
    \includegraphics[width=\linewidth]{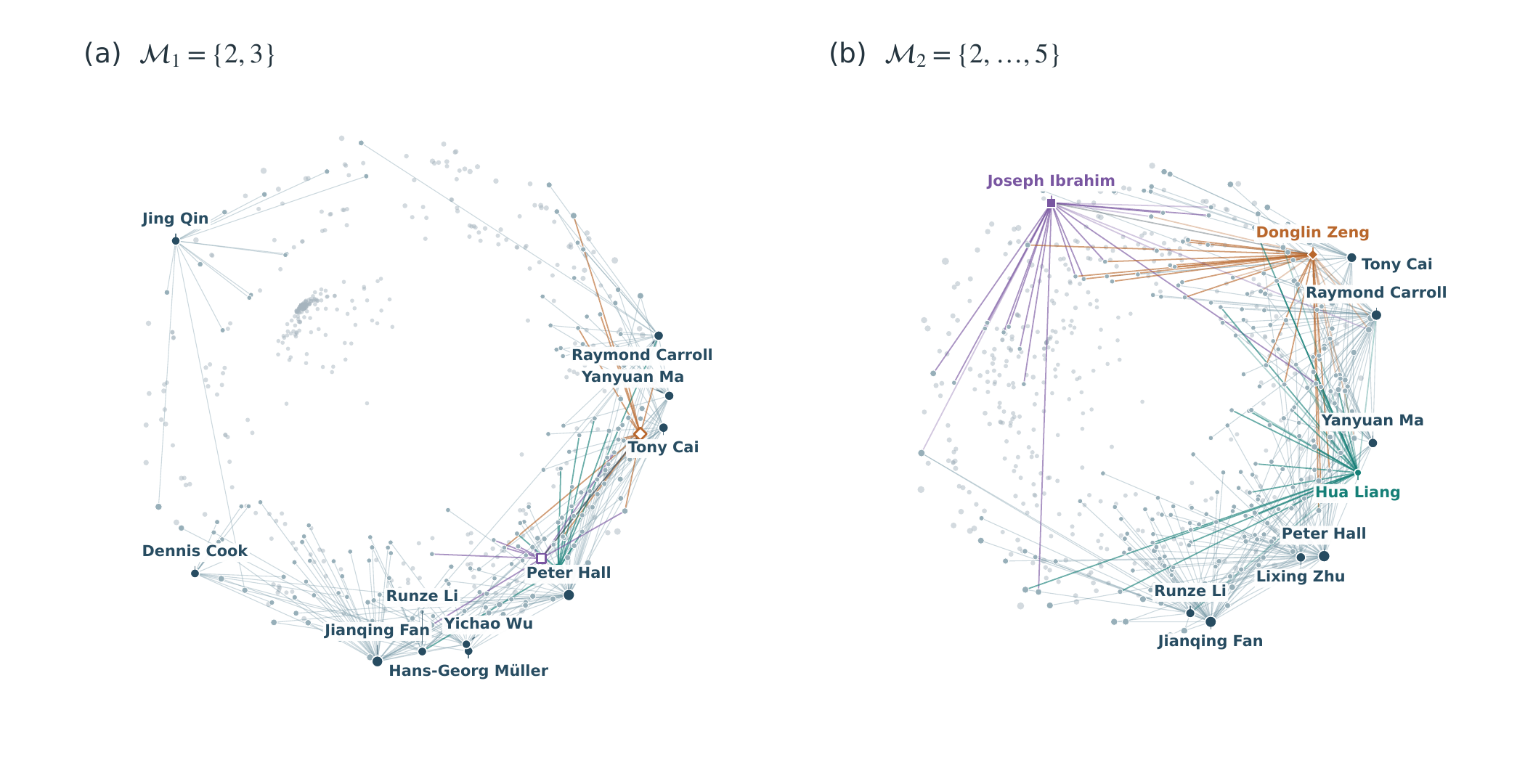}
    \caption{Hub authors in the MADStat coauthorship hypergraph from 2005 to 2015. Panels~(a) and~(b) show the latent positions of the 644 authors estimated by CINE under $\mathcal M_1=\{2,3\}$ and $\mathcal M_2=\{2,\ldots,5\}$.}
    \label{fig:real_hub}
\end{figure}

Of the 625 hyperedges, 408 involve three or more authors, including 126 with at least four authors. The observations become increasingly sparse at higher orders: the number of hyperedges decreases from 282 at order three to 89 at order four, and 20 at order five. Beyond order five, there are only eleven hyperedges of order six, three of order seven, two of order eight, and one of order ten. We therefore considered two sets of retained orders, $\mathcal M_1=\{2,3\}$ and $\mathcal M_2=\{2,\ldots,5\}$, and compared the two CINE fits through their estimated latent positions and their predictive performance on held-out data. In both fits, we set the latent dimension to $r=2$ and the sampling budget to $B_m=16|E_m^+|$ for each retained order $m>2$, a larger budget ratio than in the simulations because the MADStat hypergraph is small. Figure~\ref{fig:real_hub} displays these latent positions, which summarize patterns of coauthorship and may also reflect related research interests. For each set of retained orders, we call the ten authors who belong to the most distinct teams hubs and label them by name. Six authors are among the top ten in both fits. Donglin Zeng, Joseph Ibrahim, and Hua Liang enter the top ten only when orders four and five are retained. The numbers of distinct teams involving Zeng, Ibrahim, and Liang increase from 10 to 18, from 7 to 15, and from 9 to 18, respectively. Larger teams also reveal additional coauthorship connections for these authors. Of Zeng's 29 distinct coauthors in panel~(b), 17 do not appear with him in any two- or three-author team in our collection. Similarly, 13 of Ibrahim's 22 coauthors and 17 of Liang's 27 coauthors do not appear with them in any two- or three-author team.

\begin{figure}[htbp]
    \centering
    \includegraphics[width=\linewidth]{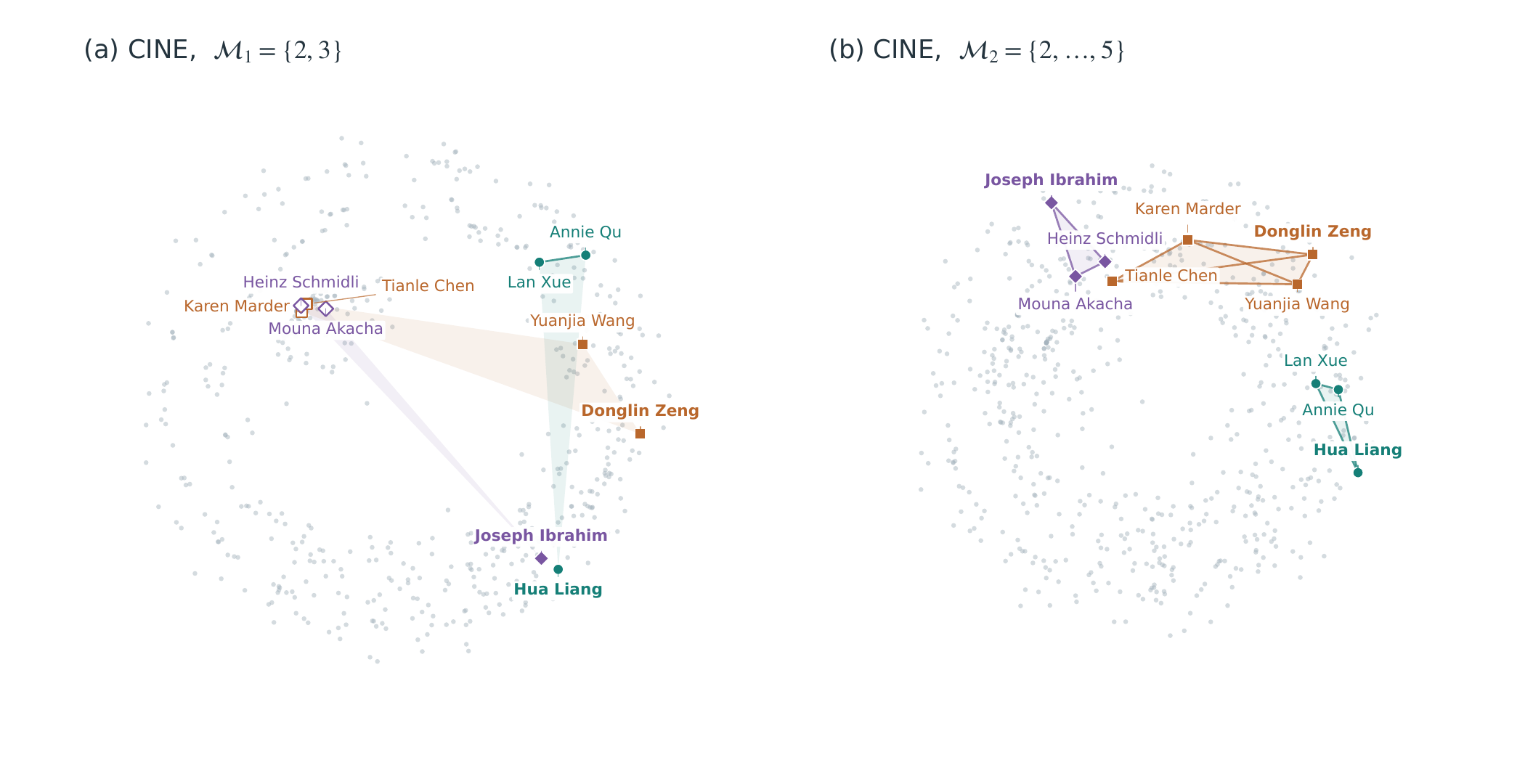}
    \caption{Estimated latent positions of authors based on the coauthorship hypergraph from 2005 to 2015. Panels~(a) and~(b) show the CINE fits under $\mathcal M_1=\{2,3\}$ and $\mathcal M_2=\{2,\ldots,5\}$.}
    \label{fig:real_group}
\end{figure}

Incorporating larger teams also changes the estimated latent positions and reveals different patterns of association. For example, Yuanjia Wang is one of Zeng's 17 additional coauthors. Figure~\ref{fig:real_group} shows that Wang lies closer to Zeng under $\mathcal M_2$ than under $\mathcal M_1$ and also moves closer to Karen Marder and Tianle Chen. All four authors collaborated on a five-author study of age-dependent risk prediction for Huntington's disease. The reduced distances from Wang to Marder and Chen are consistent with the three authors' earlier joint work outside the fitting sample. In a 2012 study, they used data from probands and their relatives to estimate the distribution of the age at onset of Huntington's disease as a function of CAG repeat length. Another example involves Joseph Ibrahim. Including larger teams brings Mouna Akacha and Heinz Schmidli closer to him in the latent space. Ibrahim, Akacha, and Schmidli collaborated with Donglin Zeng and Qingxia Chen on a five-author study of treatment effects on recurrent events in a multiple sclerosis trial. This pattern is consistent with their earlier research on correlated event times, patient dropout, and relapse rates.

A third example involves Hua Liang, Annie Qu, and Lan Xue. All three authors already appear in present teams under $\mathcal M_1$, where Qu and Xue lie close together. Under $\mathcal M_2$, the pairwise distances among these three authors all decrease, with Qu and Xue moving substantially closer to Liang. Their earlier work provides context for this pattern. Liang and Qu had previously studied variable selection from different perspectives. Liang and coauthors investigated the problem in additive partial linear models, while Qu and Cho developed penalized quadratic inference functions for correlated data. Xue had previously collaborated with both authors. She and Liang developed polynomial spline estimators for additive coefficient models in a study outside the fitting sample, while she and Qu studied model selection for correlated data. These lines of research were subsequently brought together in a four-author paper with Li Wang, which combined polynomial splines with penalized quadratic inference functions to select both linear and nonlinear effects in correlated data. Taken together, these examples suggest that larger team collaborations may reveal related research interests among statisticians that are less apparent from smaller team collaborations alone.

To assess the role of cross-order information in prediction, we compare CINE under $\mathcal M_1$ and $\mathcal M_2$ on held-out data. We assign each order-three candidate to a test set with probability $0.2$, which places 55 of the 282 order-three present hyperedges in the test set and keeps the other 227 in the training data. Let $\mathcal T_3^+$ and $\mathcal T_3^-$ denote the present hyperedges and the negative examples of order three in the test set. We measure discrimination by the area under the receiver operating characteristic curve (AUC):
\begin{equation*}
    \mathrm{AUC}=\frac{1}{|\mathcal T_3^+|\,|\mathcal T_3^-|}\sum_{e\in\mathcal T_3^+}\sum_{f\in\mathcal T_3^-}\Big[\mathbf{1}\{\widehat p_3(e)>\widehat p_3(f)\}+\tfrac{1}{2}\mathbf{1}\{\widehat p_3(e)=\widehat p_3(f)\}\Big],
\end{equation*}
where $\widehat p_3(e)$ is the fitted occurrence probability of candidate $e$. A larger AUC indicates better discrimination. CINE attains an AUC of $0.737$ under $\mathcal M_1$, and retaining orders four and five raises the AUC to $0.846$. Under $\mathcal M_2$ with the same sampling budgets, Uniform attains $0.757$, so guiding negative sampling by order-two information raises the AUC from $0.757$ to $0.846$. Cross-order information therefore improves coauthorship prediction.

Section~C of the Supplementary Material compares the latent positions and predictive performance of CINE and the alternative methods considered in the simulation study on this dataset. Overall, incorporating larger teams helped CINE reveal associations among statisticians with related research interests, and CINE also achieved the highest AUC among the methods considered under $\mathcal M_2$.
\section{Discussion}
\label{Section-Discussion}

Beyond saving computation, CINE makes higher-order information usable. In the MADStat analysis, retaining orders four and five changed the estimated latent positions. Since lower orders guide higher-order computation without changing the estimation target, users can choose the retained orders under a time budget (Remark~\ref{rem:order-selection}) and set the budget ratio to the accuracy they need, rather than being limited by the infeasibility of full estimation.

The same analysis applies to other data recording group interactions of varying size, such as communication, social contact, and biomedical co-occurrence data, especially when small groups are frequent and large groups are rare. Scores beyond pairwise affinities and samplers that grow candidates node by node for very large $n$ are natural directions for future research.

\noindent{\bf Supplementary material.} The online Supplementary Material contains the proofs and additional theoretical results (Section~\ref{Section-Theory}) and additional numerical results (Sections~\ref{Section-Simulation} and~\ref{Section-Real-Data}).

\noindent{\bf Disclosure statement.} The authors report no potential conflict of interest.

\noindent{\bf Funding.} Nie, Wang, Zhang, Zhu were partially supported by the National Natural Science Foundation of China (No.12571298), the Fundamental and Interdisciplinary Disciplines Breakthrough Plan of the Ministry of Education of China (JYB2025XDXM612), and the Fundamental Research Funds for the Central Universities.

\noindent{\bf Data and code availability.} The MADStat data \citep{ji2022co} are publicly available at \url{https://github.com/ZhengTracyKe/MADStat}. The code that reproduces the simulations and the real data analysis is provided in a zip file as part of the Supplementary Material.

\bibliography{references}

\clearpage
\includepdf[pages=-]{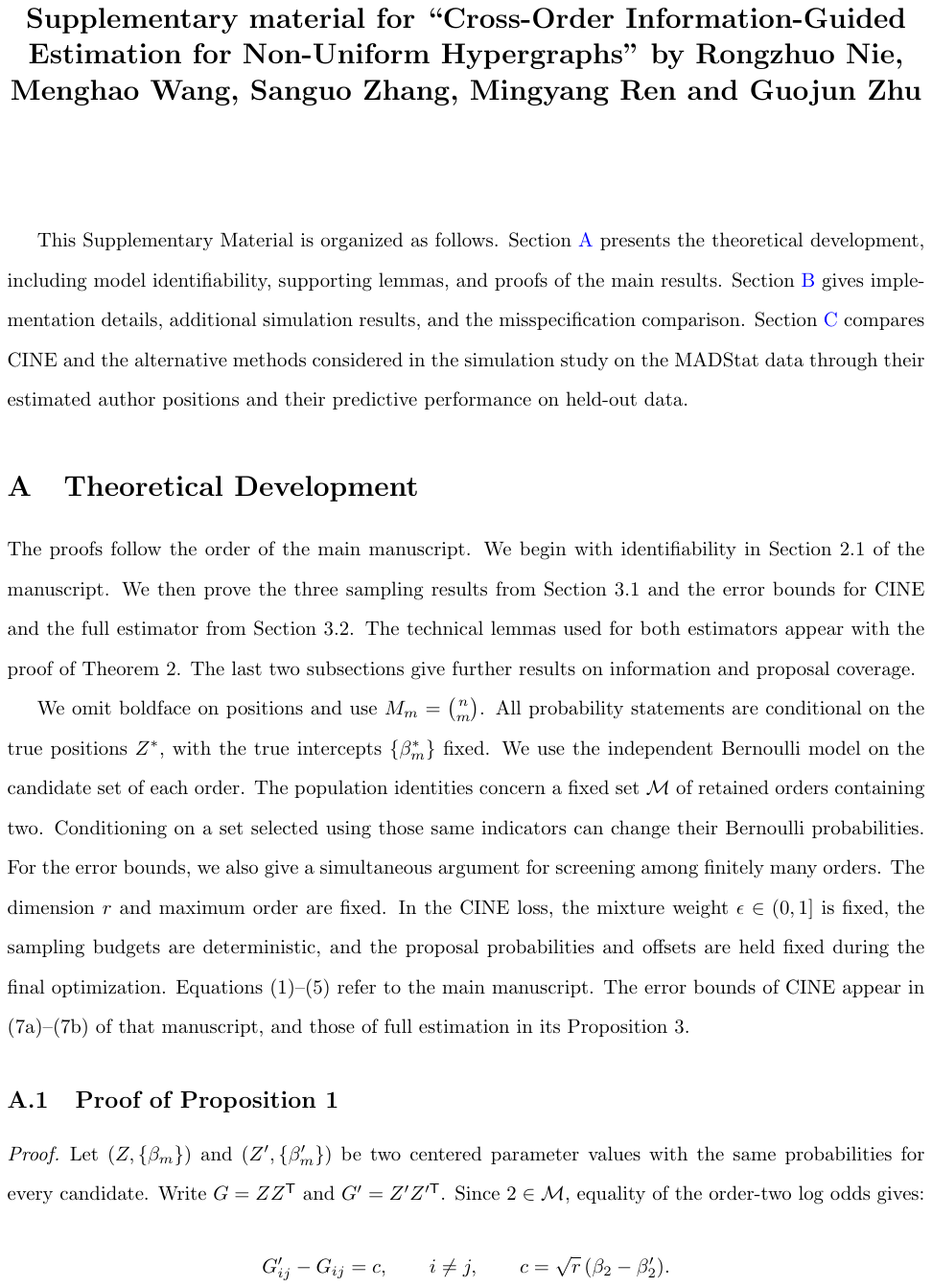}
\end{document}